\documentclass[twocolumn]{aastex6}
\usepackage{amssymb,amsfonts,amsmath,amstext,amsgen,amsopn,amsxtra,indentfirst,times}%

\usepackage{savesym}%
\usepackage{graphicx}%
\usepackage{hyperref}%
\usepackage{lineno}%
\usepackage{mathtools}%
\usepackage{multirow}
\usepackage{amsmath}%
\usepackage{xspace}%
\usepackage{units}%
\usepackage{tikz}%
\usetikzlibrary{positioning}%
\usetikzlibrary{shapes,arrows}%
\usepackage{nicefrac}%
\usepackage{array}

\usepackage{tabularx}  
\usepackage{ragged2e}  
\newcolumntype{Y}{>{\RaggedRight\arraybackslash}X} 
\usepackage{booktabs}  

\def\widesplit#1{%
\cleardoublepage
\def\row##1##2##3{##1}%
#1%
\\

\def\row##1##2##3{##2}%
#1%
\\

\def\row##1##2##3{##3}%
#1%
\clearpage
}

\newcommand{\Pbatm}{$P_{\rm batm}$\xspace}
\newcommand{\Pout}{$P_{\rm out}$\xspace}

\newcommand{\ME}{M$_{\rm \Earth}$\xspace}
\newcommand{\N}{{\texttt{N}}\xspace}
\newcommand{\U}{{\texttt{U}}\xspace}
\newcommand{\A}{{\texttt{A}}\xspace}

\newcommand{\UCM}{{\texttt{UCM}}\xspace}

\newcommand{\UHM}{{\texttt{UHM}}\xspace}
\newcommand{\UHMR}{{\texttt{UHMR}}\xspace}

\newcommand{\rcore}{$r_{\rm core}$\xspace}
\newcommand{\rsolid}{$r_{\rm core+mantle}$\xspace}

\newcommand{\prior}{{p}({\bf m})\xspace}
\newcommand{\m}{{\bf m}\xspace}
\newcommand{\dat}{{\bf d}\xspace}
\newcommand{\post}{{p}({\bf m}|{\bf d})\xspace}

\newcommand{\fesistar}{{\rm Fe}/{\rm Si}_{\rm star}\xspace}
\newcommand{\mgsistar}{{\rm Mg}/{\rm Si}_{\rm star}\xspace}
\newcommand{\fesima}{{\rm Fe}/{\rm Si}_{\rm mantle}\xspace}
\newcommand{\mgsima}{{\rm Mg}/{\rm Si}_{\rm mantle}\xspace}

\newcommand{\rev}[1]{{\color{black} {#1}}}

\begin{document}

\title{Interior characterization in multiplanetary systems: TRAPPIST-1}

\author{Caroline Dorn\altaffilmark{1}}
\author{Klaus Mosegaard \altaffilmark{2}}
\author{Simon L. Grimm\altaffilmark{1,3}}
\author{Yann Alibert \altaffilmark{3}}

\altaffiltext{1}{University of Zurich, Institut of Computational Sciences, University of Zurich, Winterthurerstrasse 190, CH-8057, Zurich, Switzerland.  Emails: cdorn@physik.uzh.ch}
\altaffiltext{2}{Niels-Bohr-Institute, Climate and Geophysics, Juliane Maries Vej 30, DK-2100 Copenhagen, Danmark}
\altaffiltext{3}{University of Bern, Center for Space and Habitability, Gesellschaftsstrasse 6, CH-3012, Bern, Switzerland.}

\begin{abstract}
Interior characterization traditionally relies on individual planetary properties, ignoring correlations between different planets of the same system. \rev{For multi-planetary systems, planetary data are generally correlated. This is because, the differential masses and radii are better constrained than absolute planetary masses and radii.
We explore such correlations and data specific to the multiplanetary-system of TRAPPIST-1  and study their value for our understanding of planet interiors.} Furthermore, we demonstrate that the rocky interior of planets in a multi-planetary system can be preferentially probed by studying the most dense planet representing a rocky interior analogue. 
Our methodology includes a Bayesian inference analysis that uses a Markov chain Monte Carlo scheme. Our interior estimates account for the anticipated variability in the compositions and layer thicknesses of core, mantle, water oceans and ice layers, and a gas envelope.
\rev{Our results show that (1) interior estimates significantly depend on available abundance proxies and (2) that the importance of inter-dependent planetary data for interior characterization is comparable to changes in data precision by 30\%.}
For the interiors of TRAPPIST-1 planets, we find that possible water mass fractions generally range from 0-25\%. The lack of a clear trend of water budgets with orbital period or planet mass challenges possible formation scenarios. \rev{While our estimates change relatively little with data precision, they critically depend on data accuracy. If planetary masses varied within $\pm$ 24\%, interiors would be consistent with uniform ($\sim$7\%) or an increasing water mass fractions with orbital period ($\sim$2-12\%).}
\end{abstract}


\section{Introduction}
\sloppy
Among all exoplanetary systems known today, the TRAPPIST-1 system harbours the largest number of Earth-sized exoplanets in a single system. It is a tightly-packed system in which  at least seven planets orbit an ultra-cool star \citep{gillon2017seven}. The proximity of the planets to their central star, the low stellar mass and stellar luminosity imply temperate conditions for the planets, with equilibrium temperatures ranging from about 160 to 400 K.  Although characteristics of the planets remind us of Earth and Venus, the characteristics of  star and  system architecture seem very exotic compared to our Solar System. How systems like TRAPPIST-1 formed and evolved over time is an open and fascinating question that has motivated plenty of exoplanetary studies.

Key to our understanding of the formation and evolution is the knowledge \rev{of the} composition and structure of the planetary interiors. Our ability to characterize planetary interiors depends on available observations and prior information.  Prior information are based on laboratory work and theoretical considerations and yield important information, e.g., on the anticipated range of possible interiors. Astrophysical observations provide, e.g., planetary mass and radius, and orbital period. 

For our study, the observational data on planetary radii and masses are taken from  \citet{delrez2018early} and \citet{Grimm2018}, respectively. The observed transit depths constrain ratios of planetary-to-stellar radii $R_{\rm p}/R_\star$ for each planet, while  the transit timing variations (TTV) constrain ratios of planetary-to-stellar masses  $M_{\rm p}/M_\star$ from gravitational interactions among all seven planets and the star. The planet bulk densities are then calculated from $\rho_{\rm p} = \rho_\star \frac{M_{\rm p}}{M_\star} \frac{R^3_\star}{R^3_{\rm p}}$, given the photometrically well-constrained stellar density $\rho_\star$ \citep{seager2003unique}. There are two important consequences that stem from this: (1) mass and radius for \emph{each} planet are correlated, and (2) planetary masses between the planets  as derived by TTV are correlated, and (3) planetary radii between the planets are correlated. 
The correlation of the masses between all planets is a consequence of the planets being in a full
resonance chain with each other. \rev{The correlation of planetary radii is because the differential  radii (or differential transit depths) are better constrained than the absolute radii because the latter includes the uncertainty on stellar radii.}
Correlation of data can have significant influences on the range of inferred    planetary interiors.

For an individual planet, the correlation of mass and radius has been taken into account for exoplanet interior characterizations \citep{weiss2016revised, crida2018mass}. However, the correlation of data between different planets of a single system has so far not been taken into account. In the present work, we develop a new resampling scheme that formally accounts for this interdependency of planetary \rev{data}.
 \rev{We demonstrate how our ability of constraining interiors is affected by data interdependencies using the example of planetary masses. We generally believe that it is important to make thorough use of \emph{all} available information (including the discussed data interdependencies) because observational data are few and expensive. Here, we provide the tool to do so. }

Previous characterization studies demonstrated the value of refractory rock-forming element abundances as constraints in addition to planetary mass and radius \citep{sotin2007mass,dorn2015can}. \rev{Unfortunately, the host star is too faint ($V=19$) to measure photospheric abundances of refractory elements with  available facilities.  Measured elemental abundances in the stellar photosphere may otherwise be used as abundance proxies for the rocky planetary interiors}. Given a multi-planetary system, we can derive an abundance proxy based on the most dense planet of the system, because its interior will be dominantly rocky.  TRAPPIST-1~e is the most dense and probably a purely rocky planet, given its high bulk density. Thus planet e may be seen as a rocky interior analogue of all planets in the system. We analyse the possible range of elemental abundances for planet TRAPPIST-1~e given only its mass and radius without abundance constraints. The obtained range of bulk abundances of TRAPPIST-1~e is subsequently used as   constraints for the other planets.  Alternatively, we also investigate the abundance constraint that was suggested by \citet{unterborn2017constraining}.

 \rev{We note that preliminary tests including new observations for TRAPPIST-1 that were previously unavailable for the study of \citet{Grimm2018} indicate shifts in planet massses within $\pm$24\% \citet{Demoryconf}. Also, \citet{kane2018impact} indicate changes in planetary radii on the order of +2\%. Therefore, our results and those of \citet{suissa2018trappist} and \citet{unterborn2018updated} should be taken with care. We show how much such systematic shifts could effect our interior estimates (Section \ref{systematics}). Nonetheless, our paper demonstrates how interior characterization for multi-planetary systems has extraordinary advances compared to individual planets.  }

The structure of our study is as follows. We briefly review previous studies on possible interiors of the TRAPPIST-1 planets (Section \ref{prv}). Then we describe our methodology that involves the new resampling scheme  (Section \ref{method}). We demonstrate and discuss our results (Section \ref{results}) in light of possible formation and evolution paths of the planets (Section \ref{discussion}), and provide a summary in the conclusions (Section \ref{conclusion}).

\section{Previous studies}
\label{prv}
Masses and radii of the planets as estimated by \citep{gillon2017seven}, suggest that all planets can have volatiles in either gas or water layers or can be rocky, while only planet f actually requires significant layer of volatiles in order to explain its low bulk density. 

Given the mass and radius estimates from \citet{gillon2017seven} and \citet{luger2017seven}, \citet{unterborn2017constraining}  estimated the water mass fraction to be more than 50 \% for planets f and g, and on the order of 7\% for planets b and c. They concluded that the difference in water mass fraction is because of a difference in formation location relative to the iceline. 

Since tidal interactions between the planets may affect the interiors, \citet{Barr2018} estimated the tidal heat fluxes on planets d, e, and f and estimated them to be twenty times higher than \rev{Earth’s} mean surface heat flow. They suggest that magma oceans could be maintained on planets b and c. 

Given the bulk densities, thick atmospheres \rev{may} cover the planets. Such thick atmospheres could be realized by few bars of hydrogen, however, atmospheric escape around TRAPPIST-1 is strong enough to limit the lifetime of H$_2$-atmospheres given the large EUV irradiation \citep{bolmont2016water} or even considering cometary impacts \cite{kral2018cometary}. Indeed, transit spectroscopy for all planets show no evidence for cloud/haze-free \rev{H$_2$}-dominated atmosphere \citep{de2016combined, de2018atmospheric}, except for planet g for which data remains inconclusive. Thus, both atmospheric observations and theoretical considerations of atmospheric loss strongly suggest the presence of terrestrial-type atmospheres.

\citet{turbet2017climate} highlight the synchronous rotation of the planets and study consequences for the diversity of climates on the planets. The nightside temperatures would allow atmospheric CO$_2$ to condense out to form ice shields. For the outer planets g and h, a global CO$_2$ ice cover is possible. However, they also show that for planets g as well as f,  greenhouse warming could be efficient enough to allow liquid water.

Loss of water over the evolution of the planets has been investigated by \citep{Bourrier_2017}. They suggest that water loss could have been efficient enough to remove few to several tens of Earth oceans over the planet's lifetime. Compared to estimates of possible water mass fractions (7-50\%), this is very small (1 Earth ocean is 0.02 \% of Earth's total mass). 

The most recent and more precise mass and density estimates from \citet{Grimm2018} provide new insights into the planet bulk compositions.
They find that purely rocky interiors are likely for planets c and e, while planets b, d, f, g, and h require envelopes of volatiles with water mass fractions generally less then about 15\% \footnote{In the published version their Figure 10 correctly compares different interiors with the planetary data. However, in the text the erroneously stated 5\% should be 15\% and 15\% should be 35 \%.}. For the inner planets b--d,  volatiles are  likely in the form of an atmosphere given the high stellar irradiation. The outer planets f--h have cold enough equilibrium temperatures such that common volatile species CO$_2$ and H$_2$O are condensed out. Planet e has been recognised by different authors \citep[e.g.,][]{turbet2017climate,Grimm2018} to have the largest potential for Earth-like surface conditions.

Here, we provide distributions of possible interiors using a Bayesian inference analysis while using the entire information from the calculated mass and radius distributions as provided by \citet{delrez2018early} and \citet{Grimm2018}. Furthermore, our physical interior model accounts for interiors of general structures and compositions that has been developed to describe the full range of super-Earths to mini-Neptunes. Also, we test different abundance proxies for refractory element and account for the correlation of planetary \rev{data} between different planets.

\begin{figure}[ht]
\centering
 \includegraphics[width = .51\textwidth, trim = 1.1cm 0cm .5cm 0cm, clip]{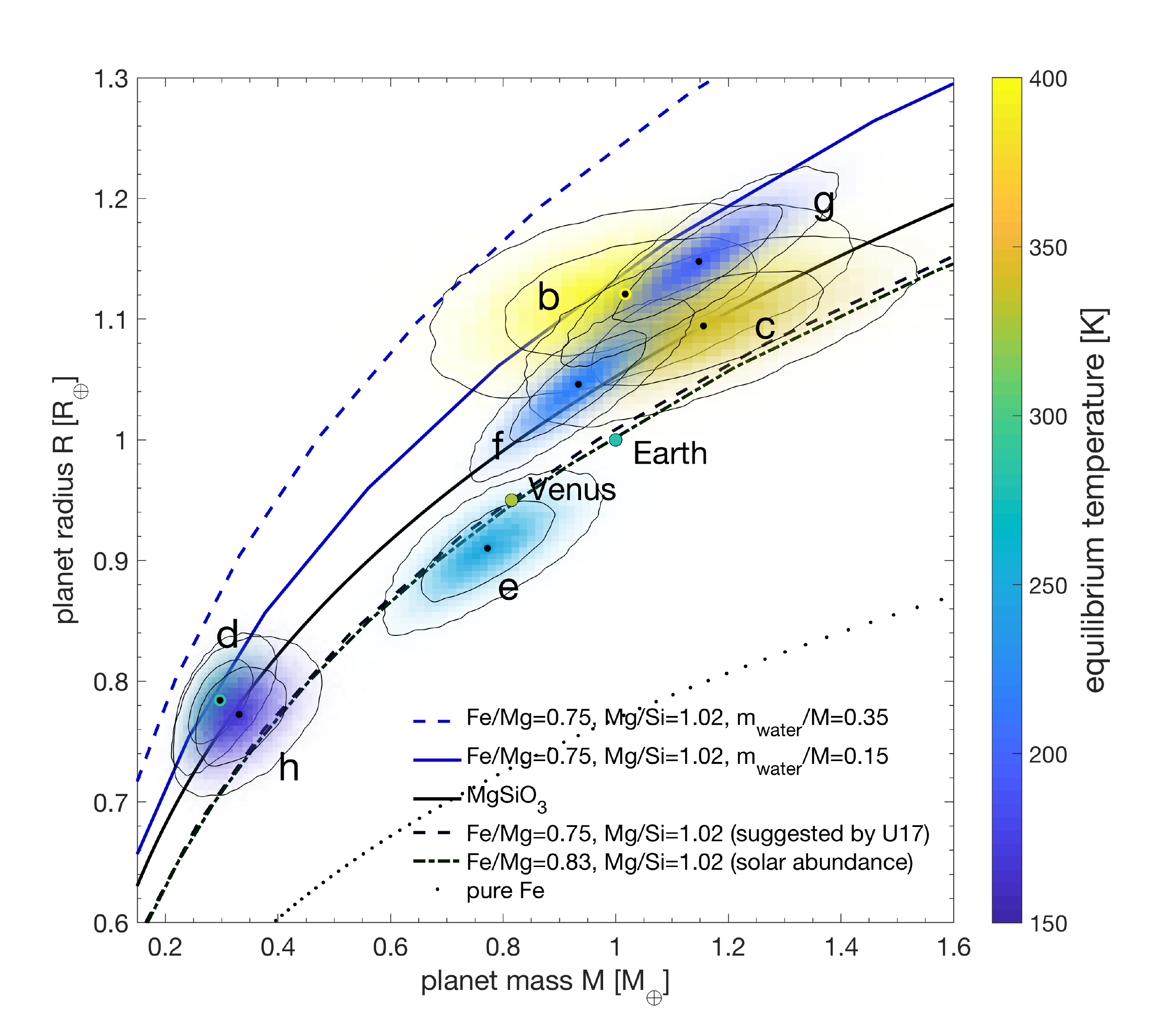}\\
 \caption{Mass-radius plot showing TRAPPIST-1 planets, Earth and Venus. Plot is adapted from \citep{Grimm2018}. U17 refers to \citet{unterborn2017constraining}.}
 \label{fig1}
\end{figure}

\begin{figure*}[ht]
\centering
 \includegraphics[width = 1.\textwidth, trim = 0cm 0cm 0cm 0cm, clip]{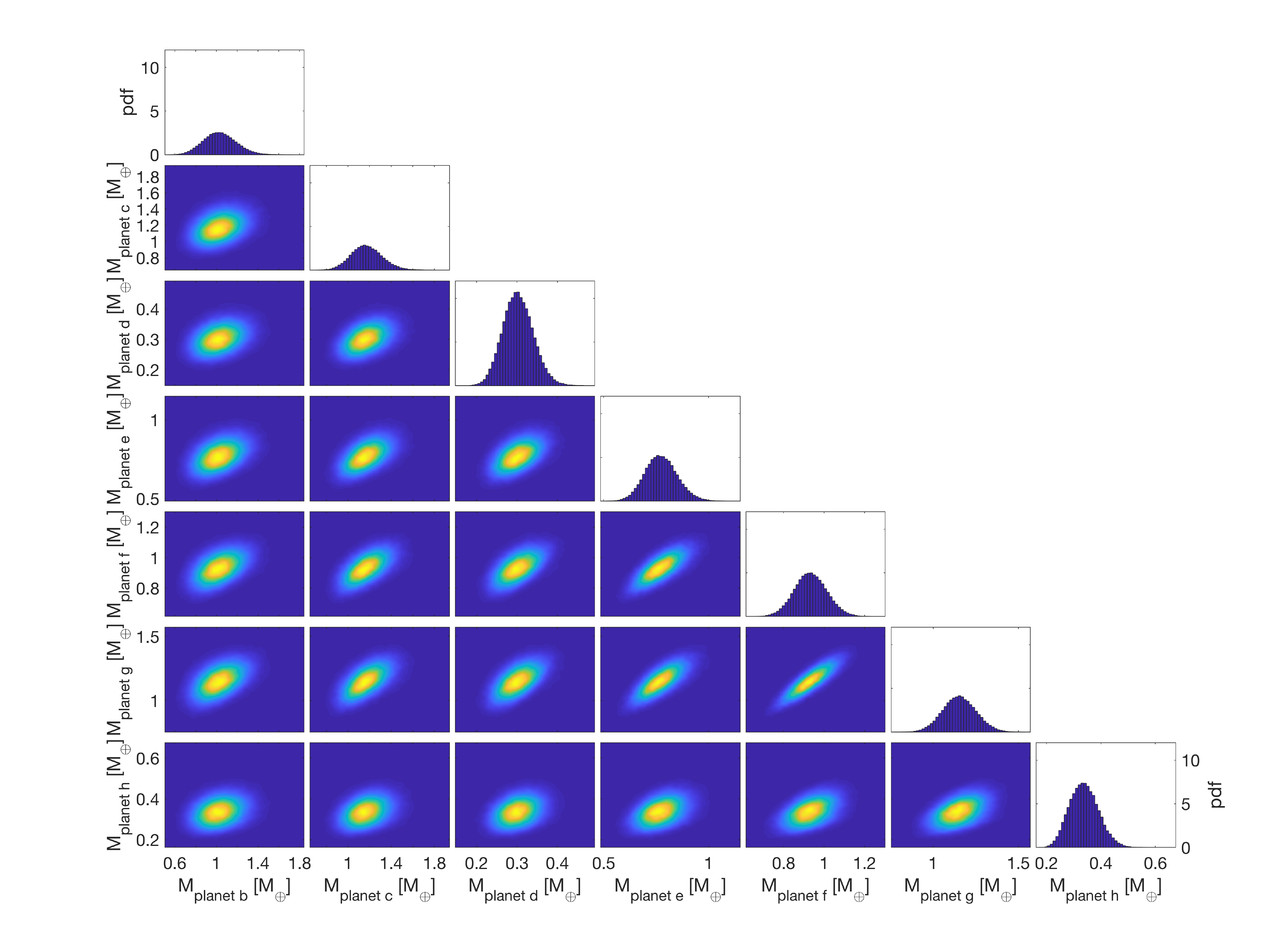}\\
 \caption{Two-dimensional marginal data distributions of planetary masses extracted from Figure 6 of \citep{Grimm2018}.}
 \label{fig:CM}
\end{figure*}

\section{Method}
\label{method}
The aim of this study is to calculate confidence regions of interior parameters that are linked to the formation and evolution of the planets. There are two sources of information that allow us to constrain interior parameters: observable data and prior information on the interior. Data \dat and interior parameters \m are linked by the interior model that is incorporated in the operator $g(\cdot)$, such that $\dat = g(\m)$.  In our case, the operator $g(\cdot)$ is not invertible. Therefore, we need to infer the possible realizations of \m given the data \dat using inference methods. Here, we use a Bayesian inference analysis based on a Markov chain Monte Carlo (McMC) scheme by \citet{dorn2015can}. This scheme characterizes every planet individually. However, there is a clear correlation of planetary data between the planets. We account for this correlation by introducing a novel resampling scheme.
In the following, we discuss the different aspects of the method.

\subsection{Interior model}

The interior model $g(\cdot)$ assumes planets to be composed of a pure iron core, a silicate mantle comprising the oxides Na$_2$O--CaO--FeO--MgO--Al$_2$O$_3$--SiO$_2$, pure water layer, and an isothermal gas layer.  Those model parameters \m that determine the rocky and the water layer are \rcore, \rsolid, $\fesima$, $\mgsima$, and $m_{\rm water}$.
 
The structural model for the interior uses self-consistent thermodynamics for core, mantle, high-pressure ice, and water ocean. 
For the core density profile, we use the equation of state (EoS) fit of iron in the hcp (hexagonal close-packed) structure provided by \citet{bouchet2013ab} on {\it ab initio} molecular dynamics simulations. 
For the silicate mantle, we compute equilibrium mineralogy and density as a function of pressure, temperature, and bulk composition by minimizing Gibbs free energy \citep{connolly2009geodynamic}. For the water layers, we follow \citet{vazan2013effect} using a quotidian equation of state (QEOS) and above 44.3 GPa, we use the tabulated EoS from \citet{seager2007mass} that is derived from DFT simulations. {Depending on pressure and temperature, the water can be in solid, liquid or super-critical phase.} Water vapour is excluded, since we impose the condition that if water is present, then there must be a gas layer on top that imposes a pressure at least as high as the vapour pressure of water.
We assume an adiabatic temperature profile within each layer of core, mantle, and solid water.
The surface temperature of the water layer is set equal to the temperature of the bottom of the gas layer. Also, temperature is assumed to be continuous across other layer boundaries.

For the gas layer, we \rev{test two different models: an isothermal (model I) and a fully adiabatic (model II). Model I assumes} a thin, isothermal atmosphere in hydrostatic equilibrium and ideal gas behavior, which is calculated using the scale-height model. Those model parameters of \m that parameterize the gas layer in model I and that we aim to constrain are the pressure at the bottom of the gas layer \Pbatm, the mean molecular weight $\mu$, and the mean temperature (parameterized by $\alpha$, see below).
The thickness of the opaque gas layer $d_{\rm atm}$ in model I is given by:

\begin{equation}
d_{\rm atm}= H \ln \frac{P_{\rm batm}}{P_{\rm out}},
\end{equation}
where the amounts of opaque scale-heights $H$ is determined by the ratio of \Pbatm and \Pout. \Pout is the pressure level at the optical photosphere for a transit geometry that we fix to 20 mbar \citep{fortney2007planetary}.
The scale-height $H$ is the increase in altitude for which the pressure drops by a factor of $e$ and can be expressed by:
\begin{equation}
H = \frac{T_{\rm atm} R^{*}}{g_{\rm batm} \mu },
\end{equation}
where $g_{\rm batm}$ and $T_{\rm atm}$ are gravity at the bottom of the atmosphere and mean atmospheric temperature, respectively. $R^{*}$ is the universal gas constant (8.3144598 J mol$^{-1}$ K$^{-1}$) and $\mu$ the mean molecular weight.
The mass of the atmosphere $m_{\rm atm}$ is directly related to the pressure $P_{\rm batm}$ as:

\begin{equation}\label{massgas}
m_{\rm atm}= 4\pi  P_{\rm batm} \frac{r_{\rm batm}^2}{g_{\rm batm}}.
\end{equation}
where $r_{\rm batm}$ is the radius at the bottom of the atmosphere, respectively.

The atmosphere's constant temperature is defined as

\begin{equation}
\label{Tequa}
T_{\rm atm}  = \alpha T_{\rm star} \sqrt{\frac{R_{\rm star}}{2 a}},
\end{equation}
where $R_{\rm star}$ and $T_{\rm star}$ are radius and effective temperature of the host star and $a$ is semi-major axes. The factor $\alpha$ accounts for possible cooling and heating of the atmosphere and can vary between 0.5 and $\alpha_{\rm max}$. The upper bound $\alpha_{\rm max}$ is because there is a physical limit to the amount of warming by greenhouse gases. We approximate $\alpha_{\rm max}$ for a moist (water-saturated) atmosphere \citep[see Appendix \ref{tlimit}, ][]{dorn2017generalized}.

The assumption of a constant mean temperature in model I might effect the interior predictions. In order to test how sensitive our results are to this assumption, we propose a second atmosphere model (model II) that assumes a convective adiabat. For model II, the temperature at the bottom of the gas layer, $T_{\rm batm} $, is calculated as:
\begin{equation}
T_{\rm batm}  = T_{\rm out} \left(\frac{P_{\rm batm}}{P_{\rm out}}\right)^{\kappa},
\end{equation}
where $T_{\rm out}$ is calculated with Equation \ref{Tequa}, while restricting $\alpha$ to values between 0.5 and 1, which is equivalent to a range of albedos from 0 to 0.94. The exponent $\kappa$ is equal to $2/(2+n)$, with $n$ being the number of degrees of freedom (which we  vary randomly between 5 and 6 for diatomic or triatomic gases, thus neglecting vibrational modes). 
Following an convective adiabat, the thickness of the opaque atmosphere is set to:
\begin{equation}
d_{\rm atm}= \frac{c_p}{g_{\rm batm}} (T_{\rm batm} -T_{\rm atm}),
\end{equation}
where the heat capacity is $c_p = (n/2 + 1) R^{*}/\mu$.

In summary, the entire interiors are parameterized by the full set of interior parameters \m: \rcore, \rsolid, $\fesima$, $\mgsima$, $m_{\rm water}$, \Pbatm, $\alpha$, $\mu$ and  for model II also $n$.
We refer to \citet{dorn2017generalized} for more
details on the interior structure model.

\subsection{Prior information}

The prior distributions of the interior parameters are listed in Table \ref{tableprior}. The priors are chosen conservatively. The cubic uniform priors on \rcore and \rsolid reflect equal weighing of masses for both core and mantle. Prior bounds on $\fesima$ and $\mgsima$ are determined by the abundance proxies. Since iron is distributed between core and mantle, the abundance constraint only sets an upper bound on  $\fesima$ thus making iron-free mantles possible. 
For the atmospheric models I \& II, a ln-uniform prior is set for \Pbatm. P$_{\rm batm, max}$ is estimated by Equation 24 from \citet{ginzburg2016super} who investigated Super-Earth atmospheres. A uniform prior in $\mu$ equally favors various dominating species (e.g., N$_2$, CO$_2,$ H$_2$O), which seems appropriate for anticipated terrestrial-like atmospheres. The factor $\alpha$ incorporates possible cooling (model I \& II) and heating  (model I) of the atmosphere, it can vary linearly between 0.5 and $\alpha_{\rm max}$ for the isothermal model I, and between 0.5 and 1 for the adiabatic model II. The upper bound $\alpha_{\rm max}$ is a physical limit to the amount of warming by greenhouse gases and is described in Appendix \ref{tlimit}. The lower bound of 0.5 is equivalent to an albedo of 0.94, which is similar to the highest albedo estimate among Solar System bodies (0.96 for Eris, \citet{sicardy2011pluto}).

\begin{table}[ht]
\caption{Prior ranges.  \label{tableprior}}
\begin{center}
\begin{tabular}{lll}
\hline\noalign{\smallskip}
parameter & prior range & distribution  \\
\noalign{\smallskip}
\hline\noalign{\smallskip}
\rcore         & (0.01  -- 1) $r_{\rm mantle}$ &uniform in $r_{\rm core}^3$\\
$\fesima$           & 0 -- $\fesistar$&uniform\\
$\mgsima$         & $\mgsistar$ &Gaussian\\
\rsolid   & (0.01 -- 1) $R$& uniform in $r_{\rm mantle}^3$\\
$m_{\rm water}$ & 0 -- 0.98 $M$& uniform\\
\Pbatm  & 20 mbar -- P$_{\rm batm, max}$ & uniform in $\ln$(\Pbatm)\\
$\alpha$  (model I)           & 0.5 -- $\alpha_{\rm max}$ &uniform \\
$\alpha$  (model II)                  & 0.5 -- 1. &uniform \\
$\mu$                 & 2.3 -- 50.0& uniform in 1/$\mu$\\
$n$ (model II) & 5--6& uniform\\
\hline
\end{tabular} 
\end{center}
\end{table}

\subsection{Data}

The data that we use to characterize the interiors of the planets comprise planetary mass and radius, stellar irradiation, and constraints on refractory element abundances. 
Planetary radii are taken from \citet{delrez2018early}; masses and semi-major axes are taken from  \citet{Grimm2018}, while stellar radius and luminosity are given by \citet{gillon2017seven}.
These data are based on several hundred planet transits  that were observed between September 2015 and March 2017. In addition to the transits presented in earlier studies \citep{gillon2016,gillon2017seven,de2016combined, luger2017seven},  \citet{Grimm2018}  included additional observations from \emph{Spitzer Space Telescope}, \emph{Kepler}, and \emph{K2}.

 For TRAPPIST-1, there is a correlation between planetary radius and mass for the individual planets as well as a correlation between all seven planetary masses and between all seven planetary radii. 
In order to better distinguish between the two kinds of correlation we will use the following nomenclature.
Unless otherwise mentioned, the terms \emph{correlated data} and \emph{uncorrelated data} refer to the correlation between mass and radius of individual planets, while the terms \emph{dependent} and \emph{independent} refer to the correlation between planetary data of the different planets.
Planetary masses and radii are  shown in Figure \ref{fig1} and listed in Table \ref{tab:mr}, where we also list the corresponding correlation $c_{\rm m,r}$. Figure \ref{fig:CM} shows the interdependency of planetary masses as derived from \citet{Grimm2018}.

For the constraints on refractory element abundances we have compiled two different scenarios. In the first scenario (\U), we use the bulk abundance constraints as suggested by \citet{unterborn2017constraining}, who analysed F-G-K stars of similar metallicity to TRAPPIST-1 and estimated their statistical median to Fe/Mg$ =  1.72 \pm 0.46$ in mass ratios, while fixing the Mg/Si mass ratio to 0.87. Here we adopted these estimates: Fe/Si $= 1.49 \pm 0.4$, Mg/Si $= 0.89 \pm 0.3$. For comparison, the solar mass ratio estimates are Fe/Si = 1.69 and Mg/Si = 0.89 \citep{lodders20094}.
In a second scenario (\A), we do not impose any constraints on bulk abundances that are based on stellar estimates. Instead, we analyse the possible range of refractory elements for the most dense planet, TRAPPIST-1~e, while only using its data of mass, radius, and stellar irradiation. The estimated range of Mg/Si$_{\rm T1e}$ and Fe/Si$_{\rm T1e}$ are subsequently used as input constraints under the premise that planet e is analogous in its relative refractory element budget to the rocky interiors of all planets. This excludes \emph{a priori} the possibility of planet e being a Mercury-type planet.

In total, we discuss five different data scenarios, labelled \U, \A, \UCM, \UHM, \UHMR, that we analyse in detail. All data scenarios are summarized in Table \ref{tabledata}. Scenarios \U and \A differ in terms of the abundance constraints used; in scenario \UCM, we additionally account for fact that planetary masses of different planets depend on each other; for \UHM and \UHMR we use hypothetical reduced uncertainties on planetary masses and radii. All scenarios help to determine the value of the different data with respect to making interior predictions.

\begin{table*}
  \caption{Masses, radii and their correlation coefficients $c_{\rm m,r}$ of all 7 planets.}
 \label{tab:mr}
  \centering
    \begin{tabular}{c | c c c | c c c | c}
planet & m [$M_{\oplus}$] & $-\sigma$ & +$\sigma$ &R [$R_{\oplus}$] & $-\sigma$ & +$\sigma$ & $c_{\rm m,r}$\\
\hline
b&1.017&0.143&0.154&1.121&0.032&0.031&0.502\\
c&1.156&0.131&0.142&1.095&0.031&0.030&0.624\\
d&0.297&0.035&0.039&0.784&0.023&0.023&0.569\\
e&0.772&0.075&0.079&0.910&0.027&0.026&0.708\\
f&0.934&0.078&0.080&1.046&0.030&0.029&0.855\\
g&1.148&0.095&0.098&1.148&0.033&0.032&0.863\\
h&0.331&0.049&0.056&0.773&0.027&0.026&0.386\\
  \end{tabular}
\end{table*}

\begin{table*}[ht]
\caption{Data scenarios for interior characterization.  \label{tabledata}}
\begin{center}
\begin{tabular}{l|llllllll}
\hline\noalign{\smallskip}
considered data & \N &\U&  \A& \UCM  & \UHM & \UHMR \\
\noalign{\smallskip}
\hline\noalign{\smallskip}
correlated $M_{\rm p}$ \& $R_{\rm p}$ & \checkmark& \checkmark& \checkmark &  {\checkmark}   & \checkmark & \checkmark \\
stellar irradiation & \checkmark& \checkmark & \checkmark &  \checkmark   &\checkmark & \checkmark \\
interdependency of planetary masses $M_{\rm p}$ & $\times$& $\times$ & $\times$ &   \checkmark  &$\times$ & $\times$ \\
interdependency of planetary masses $M_{\rm p}$ excl. planet e & $\times$& $\times$ & $\times$ &   $\times$ &  $\times$ & $\times$ \\
bulk abundances based on \citep{unterborn2017constraining} & $\times$& \checkmark  &  $\times$  & \checkmark&  \checkmark&  \checkmark \\
bulk abundances based on TRAPPIST-1~e & $\times$& $\times$  &  \checkmark  &$\times$ &$\times$  & $\times$  \\
hypothetical reduced precision in $M_{\rm p}$ & $\times$& $\times$ & $\times$ &  $\times$ & $\times$  &\checkmark  \\
hypothetical reduced precision in $R_{\rm p}$ & $\times$& $\times$ & $\times$ &  $\times$ &\checkmark & \checkmark \\
\hline
\end{tabular} 
\end{center}
\end{table*}



\subsection{Interior characterization scheme}

Our characterization scheme involves two main steps. The first step comprises the interior characterization of individual planets b--h using the generalized Bayesian inference scheme of \citet{dorn2017generalized}. The output are posterior distributions for each of the planets that do not depend on each other. In a second step, we account for the interdependency of the different planetary data. In order to do so, we have developed a resampling scheme which yields posterior distributions for each of the planets that depend on each other. Thereby, we take \emph{all} available information of the mass-radius-data into account. 

\begin{figure}[ht]
\centering
\includegraphics[width = .5\textwidth, trim = 6.3cm 2.1cm 4.1cm 2.5cm, clip]{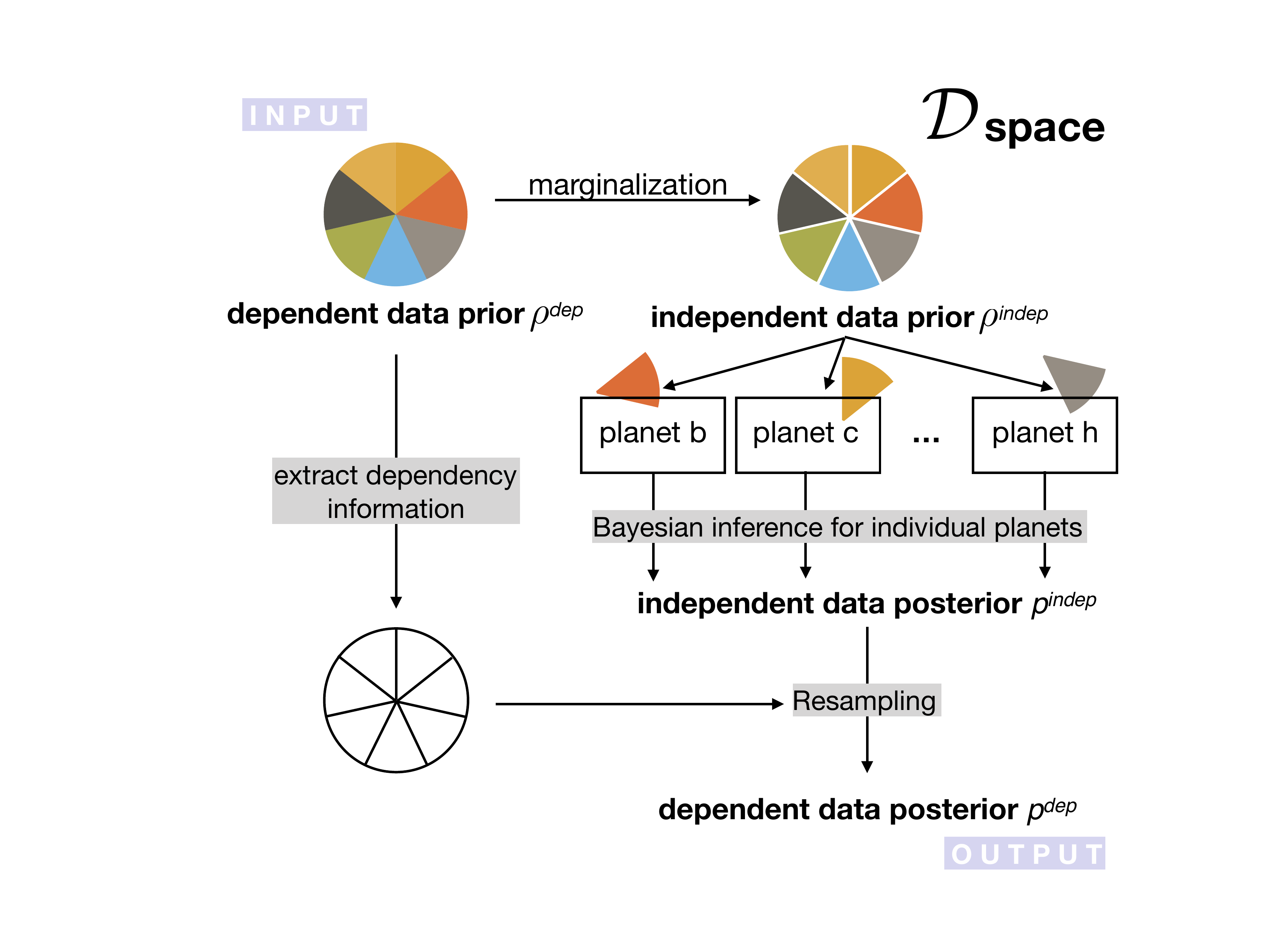}\\
 \caption{Schematic of interior characterization scheme represented in the data space $\mathcal{D}$. \emph{Dependent data} refers to the data that contains the interdependency of different planetary \rev{data}. \emph{Independent} data refers to the data of each planet that do not depend on other planets.}
 \label{scheme1}
\end{figure}

In the following, we first introduce the formulation of \citet{tarantola1982generalized} on inference problems, and second we describe each of the two main steps. Our scheme is summarized in Figure \ref{scheme1}.

\subsubsection{Bayes' theorem and the Tarantola-Valette formulation}

Formally, the posterior distribution based on Bayes' theorem for a fixed model parameterization and conditional on data is given by:
\begin{equation}
\post = \frac{\prior L({\bf m}|{\bf d})}{p(\dat)} \,
\end{equation}
where $\prior$ represents prior information on model parameter \m and $L({\bf m}|{\bf d})$ is the likelihood function, and $ p(\dat)$ signifies the evidence and is here a normalizing constant. The likelihood function is defined as:
\begin{equation}
\label{like}
\begin{split}
L({\bf m}|{\bf d}) &= \frac{1}{(2\pi)^{N/2} (\det \Sigma)^{1/2}}\\
 &\times \exp \left(- \frac{1}{2} (g(\m) - \dat)^T \Sigma^{-1} (g(\m)-\dat)\right) \,,
\end{split}
\end{equation}
where $\Sigma$ is the data covariance matrix and $\det(\Sigma)$ denotes the determinant of $\Sigma$. For uncorrelated data errors, $\Sigma$ is a diagonal matrix. Here, the correlation between planetary mass and radius is prescribed by the off-diagonal entries of $\Sigma$.

There is a different formulation of inference problems by \citet{tarantola1982generalized} that we will use in the following.
In this formulation, the posterior $p(\dat,\m)$ is expressed in the joint space of data and model parameters ($\mathcal{D}$ and $\mathcal{M}$) as:

\begin{equation}
p(\dat,\m) \equiv (\rho \land \theta)(\dat,\m) = \frac{\rho(\dat,\m) \theta(\dat,\m) }{z(\dat,\m)} \,
\end{equation}
where the prior $\rho$, the forward model density $\theta$, and the null-information density $z$ are also defined in the joint space $\mathcal{D} \times \mathcal{M}$.  The null-information is a normalizing constant. The posterior of the model parameters evaluated in the model space is then:
 \begin{equation}
p_m(\m) = \int_\mathcal{D}{p(\dat,\m) d\dat}
\end{equation}
In Appendix \ref{BayesTV}, we show that the Tarantola-Valette formulation is equivalent to Bayes' theorem under simplified conditions.

\subsubsection{Bayesian inference of individual planets}

We use the probabilistic inference analysis of \citep{dorn2017generalized} who employ a Markov chain Monte Carlo method (McMC). This method is used for every planet individually and it computes posterior probability density function (pdf) for each interior parameter \m from data \dat and prior information. It is important to note the data \dat used for each planet characterization is the marginalized data for each planet, such that data priors $\rho(\dat,\m)$ are independent. Thus, the obtained posteriors are independent of each other. The information on interdependency of data from different planets is ignored at this step and instead taken into account by our developed resampling scheme.  
The Bayesian inference analysis, in essence, explores the model space by sampling a huge number of model realizations \m that are distributed
according to the posterior distribution $p(\dat,\m)$. This is achieved within an iterative scheme where the ratio of likelihoods (Eq. \ref{like}) between subsequent sampled models is used as a criteria to accept or reject model realizations. The likelihood in probabilistic terms is a measure of how well a model realization fits the observed data.
Here, the posterior information is gathered from a large number of sampled models ($\sim 10^6$). We refer to \citep{dorn2015can} for further details on the inference analysis.

\subsubsection{Resampling scheme to account for the interdependency of data  from different planets}
\label{sec:resamp}
The above described Bayesian inference analysis yields posterior samples in the model and data space ($\mathcal{M}$ and $\mathcal{D}$), p(\dat,\m). 
Note that the posterior data  $p_d(\dat)$ can be different from the prior data $\rho_d(\dat)$. The data prior is the distribution of the noise-contaminated data which is inherited by the given distribution of observational uncertainties. The posterior data is recomputed data, taking into account constraints from the model \m. Prior data and posterior data are therefore not neccessarily identical. The aim of our resampling is to preserve the differences between posterior and prior (that is added by model \m), while adding the information of the \emph{interdependency} of prior data but not the prior data itself (since these have been already used in the Bayesian inference).

The posteriors $p(\dat,\m)$ are obtained for each planet individually. Consequently, the obtained data posteriors for a specific planet evaluated in $\mathcal{D}$, $p_{d}(\dat)$, does not depend on the posteriors of other planets, but they are independent. The interdependency of planetary masses as described by \citet{Grimm2018} quantifies the dependencies of data priors between the planets. Formally, we can say that $\rho_{d_{ij}}(\dat_{ij}) \neq \rho_{d_i}(\dat_i) \rho_{d_j}(\dat_j)$, where $i,j$ are generic planet indices $i,j \in \{b,c,d,e,f,g,h\}$.
In case the individual data priors are independent from each other, we can write the data posteriors as:
 \begin{equation}
p_{d_{ij}}^{indep}(\dat) = L_{m_i}(\dat_i)L_{m_j}(\dat_{j}) \rho_{d_i}(\dat_i)\rho_{d_j}(\dat_j) \,
\end{equation}
while for dependent data priors, the posteriors are:
 \begin{equation}
\begin{split}
p_{d_{ij}}^{dep}(\dat) & = L_{m_{ij}}(\dat_i,\dat_{j}) \rho_{d_{ij}}(\dat_i,\dat_j)  \\
                                   & = L_{m_i}(\dat_i)L_{m_j}(\dat_{j}) \rho_{d_{ij}}(\dat_i,\dat_j)   
\end{split}
\end{equation}

If $L_{m_{ij}}(\dat_i,\dat_{j}) = L_{m_i}(\dat_i)L_{m_j}(\dat_{j}) $, then
 \begin{equation}
\label{eq_res}
p_{d_{ij}}^{dep}(\dat_{ij}) = p_{d_{ij}}^{indep}(\dat_{ij}) \frac{ \rho_{d_{ij}}(\dat_i,\dat_j) }{ \rho_{d_i}(\dat_i)\rho_{d_j}(\dat_j)}
\end{equation}

The condition that $L_{m_{ij}}(\dat_i,\dat_{j}) = L_{m_i}(\dat_i)L_{m_j}(\dat_{j}) $ can be understood from the following.
$L({\bf d})$ is a measure of the likelihood of the specific dataset ${\bf d}$, based solely on the prior information about the models ${\bf m} = g^{-1}({\bf d})$ that map into ${\bf d}$. If $g$ is bijective (the inverse problem has a unique best-fitting solution), $L_{m}({\bf d}) = \rho_m (g^{-1}({\bf d}))$. If the problem is non-unique, the right-hand-side of this equation expresses an integral/sum over all models that map into ${\bf d}$. For simplicity, consider a case where our data consist of two (possibly dependent) parts, ${\bf d}_1$ and ${\bf d}_2$. In this case, the data likelihood can be expressed as
\begin{equation}
L_{m}({\bf d}) = L_{m}({\bf d}_1, {\bf d}_2) =\rho_{m_1,m_2} (g^{-1}({\bf d}_1, {\bf d}_2))
\end{equation}
However, when ${\bf m}_1$ and ${\bf m}_2$ are a priori (statistically) independent, and ${\bf m}_i$ is functionally related only to ${\bf d}_i$  (for $i = 1,2$), we get
\begin{align}
\rho_{m_1,m_2} (g^{-1}({\bf d}_1, {\bf d}_2)) &= 
\rho_{m_1} (g^{-1}({\bf d}_1, {\bf d}_2)) \cdot
\rho_{m_2} (g^{-1}({\bf d}_1, {\bf d}_2)) \\
&= \rho_{m_1} (g^{-1}({\bf d}_1)) \cdot
\rho_{m_2} (g^{-1}({\bf d}_2))
\end{align}
leading to
\begin{equation}
L_{m}({\bf d}_1, {\bf d}_2) = L_{m_1}({\bf d}_1) \cdot L_{m_2}({\bf d}_2) \ .
\end{equation}
So far for the theory.

In practice, the resampling works as follows. The inference analysis of individual planets yields independent data posteriors $p_{d_{ij}}^{(indep)}(\dat_i,\dat_j) = p_{d_i}(\dat_i) p_{d_j}(\dat_j)$. For a total of $K$ samples, we have $d_i^{1}$, $d_i^{2}$, ..., $d_i^{K}$ from $p_{d_i}$ and $d_j^{1}$, $d_j^{2}$, ..., $d_j^{K}$ from $p_{d_j}$. Thus, any pair (e.g., $(d_i^{1},d_j^{1})$, $(d_i^{2},d_j^{2})$, ... $(d_i^{K},d_j^{K})$) is a sample from $p_{d_i}(\dat_i) p_{d_j}(\dat_j)$.
We shall resample from $p_{d_i}(\dat_i) p_{d_j}(\dat_j)$ such that the new samples are drawn from $p_{d_{ij}}^{dep}(\dat_i,\dat_j)$. In order to do so, we calculate a factor $Q$ that signifies a maximum difference between the data priors between the data-independent and the data-dependent case:
\begin{equation}
Q = \max_k ( \rho_{d_{ij}}^{dep}(\dat_i^k,\dat_j^k) /  \rho_{d_i}(\dat_i^k) \rho_{d_j}(\dat_j^k)) \,.
\end{equation}

For $k = 1,...,K$:
\begin{enumerate}
\item we generate a random number $w \in [0,1]$
\item if $w<  \rho_{d_{ij}}^{dep}(\dat_i^k,\dat_j^k) /  (Q \times \rho_{d_i}(\dat_i^k) \rho_{d_j}(\dat_j^k))$, then $(\dat_i^{k},\dat_j^{k}) = (\dat_i^{k},\dat_j^{k})^{dep}$, otherwise discard the sample.
\end{enumerate}
Thereby, we obtain samples $(\dat_i^{k},\dat_j^{k})^{dep}$ that represent the data posterior $p_{d_{ij}}^{dep}(\dat_i,\dat_j)$. Any interior model pairs $(\m_i^{k},\m_j^{k})^{dep}$ that generated $(\dat_i^{k},\dat_j^{k})^{dep}$ are samples from the model posterior $p_{m_{ij}}^{dep}(\m_i,\m_j)$.

In words, we can understand the resampling as follows. 
We randomly combine samples from the analysis of individual planets i and j, which represents a set of paired samples $(d_i^{k},d_j^{k})$ drawn from the data posterior under the premise that data of different planets are independent. In order to obtain the data posterior samples $(d_i^{k},d_j^{k})^{dep}$ that incorporate the dependencies of data from different planets, we reject some of the original samples $(d_i^{k},d_j^{k})$. Those samples for which the ratio between  dependent data prior and  independent data prior are large (close to one or larger), are likely accepted as samples from $(d_i^{k},d_j^{k})^{dep}$, while they are rejected if the ratio is closer to zero.
This procedure is done for all combinations of planets $i$ and $j$.

Alternatively, it is in principle possible to do an inference analysis by using all data and their correlations and interdependencies between planets. However, our scheme elegantly separates the information of correlated mass and radius of \emph{individual} planets from information about interdependent data of \emph{different} planets. This allows us to better control each analysis steps and to be more flexible with testing different data scenarios without the need to run a full inference analysis for the entire system. Furthermore, our scheme can be generally applied to other inference problems.

\section{Results}
\label{results}

\subsection{Bulk refractory element constraints}
\label{abuabu}

\begin{figure}[ht]
\centering
\includegraphics[width = .5\textwidth, trim = 0cm 0cm 0cm 0cm, clip]{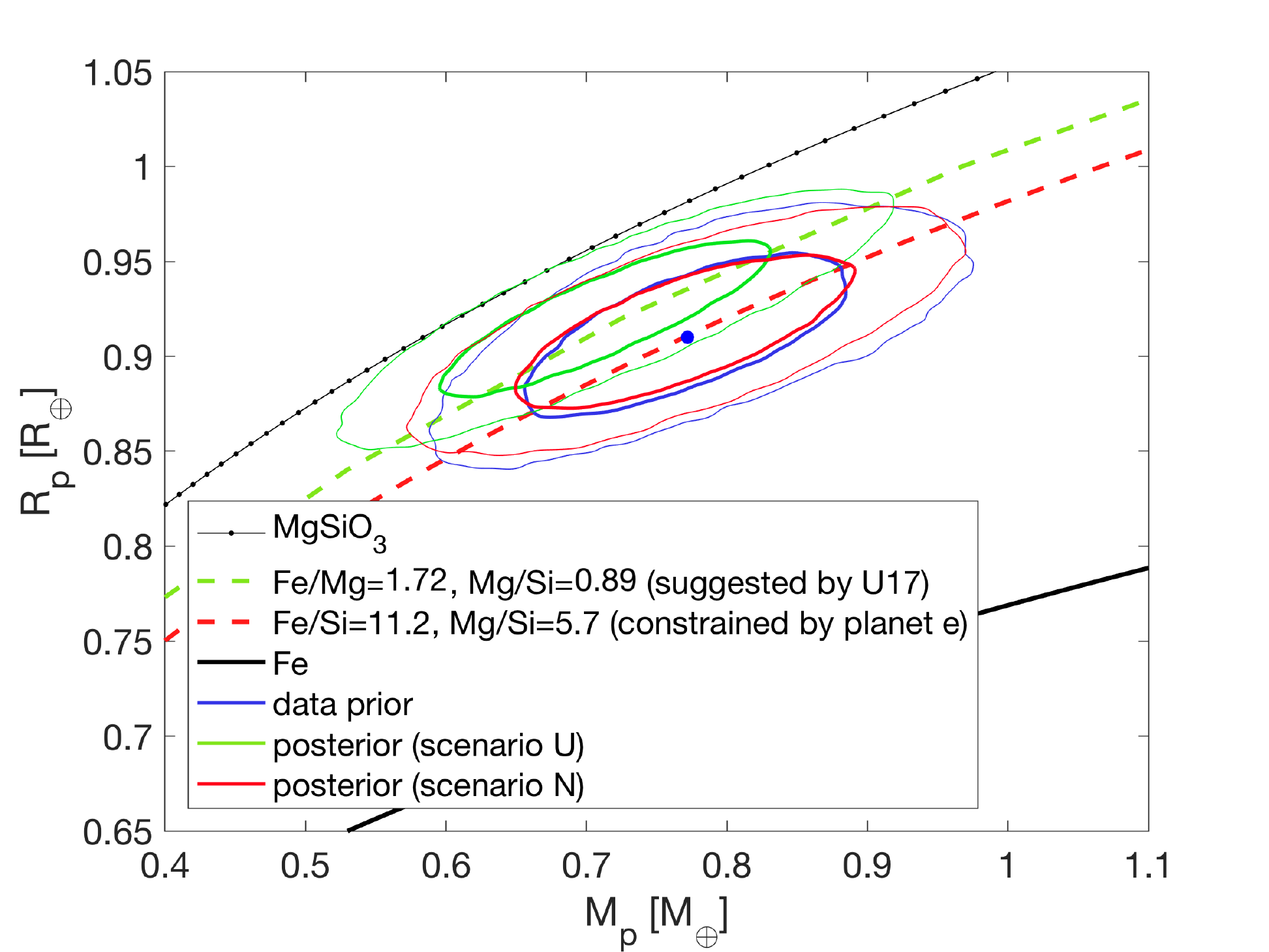}\\
 \caption{Comparison of data priors and data posteriors for TRAPPIST-1~e for planetary mass and radius. The prior distribution \citep{Grimm2018} is shown in blue which overlaps well with the posterior distribution of scenario \N, but not only in parts with the posterior distribution of scenario \U. In comparison different mass-radius curves are plotted for: pure iron (black solid), pure MgSiO$_3$ (black point-solid), interiors of iron-free mantles that agree with the median constraint of \citet{unterborn2017constraining} (green dashed), and interiors of median structure and composition as inferred for planet e and scenario \N (red dashed). Stated ratios are mass ratios.}
 \label{Fig:T1e}
\end{figure}

In Figure \ref{Fig:T1e}, we show the consequences of different bulk constraints on mass and radius of TRAPPIST-1~e. The prior data of measured mass and radius is highlighted in blue. The use of the suggested stellar constraint of  \citet{unterborn2017constraining} (scenario \U) is not able to explain well the relatively high bulk density of TRAPPIST-1e. This is evident from the difference between the prior data distribution (blue) and the posterior distribution of scenario \U (green). 

In a second scenario \N, we do not impose any abundance constraints. Instead, we analyse the possible range of bulk abundances of TRAPPIST-1~e, while only using the data of mass, radius, and stellar irradiation. The predicted bulk mass ratios are high with large 1-$\sigma$ uncertainties: Fe/Si$_{\rm T1e}= 11.2 \pm 5.7$ and Mg/Si$_{\rm T1e} = 5.7 \pm 3.7$.  The data of only mass and radius (\N) allows high relative bulk iron abundances, since iron incorporated in silicates and oxides in the mantle have a limited influence on bulk density.
These abundance ranges (Fe/Si$_{\rm T1e}$  and Mg/Si$_{\rm T1e}$) are subsequently used as input constraints to analyse  the other planets in scenario \A.

The ratios Fe/Si$_{\rm T1e}$ and Mg/Si$_{\rm T1e}$ are much higher than those estimated for Earth and other Solar terrestrial planets. However, the large uncertainties on the ratios also allow for Earth-like rocky compositions. The fact that these uncertainties estimated from mass and radius alone are high, illustrates the large degeneracy of rocky interiors, i.e., very different combinations of mantle compositions and core sizes can explain a given planetary mass and radius. The suggested stellar abundance proxy of \citet{unterborn2017constraining} with a relatively small uncertainty of  30 \%, is partly incompatible with mass and radius of TRAPPIST-1~e. In the discussion (Section \ref{discussion}), we provide two interpretations for the interior of planet e.

\begin{figure*}[ht]
\centering
\includegraphics[width = .8\textwidth, trim = 0cm 0cm 0cm 0cm, clip]{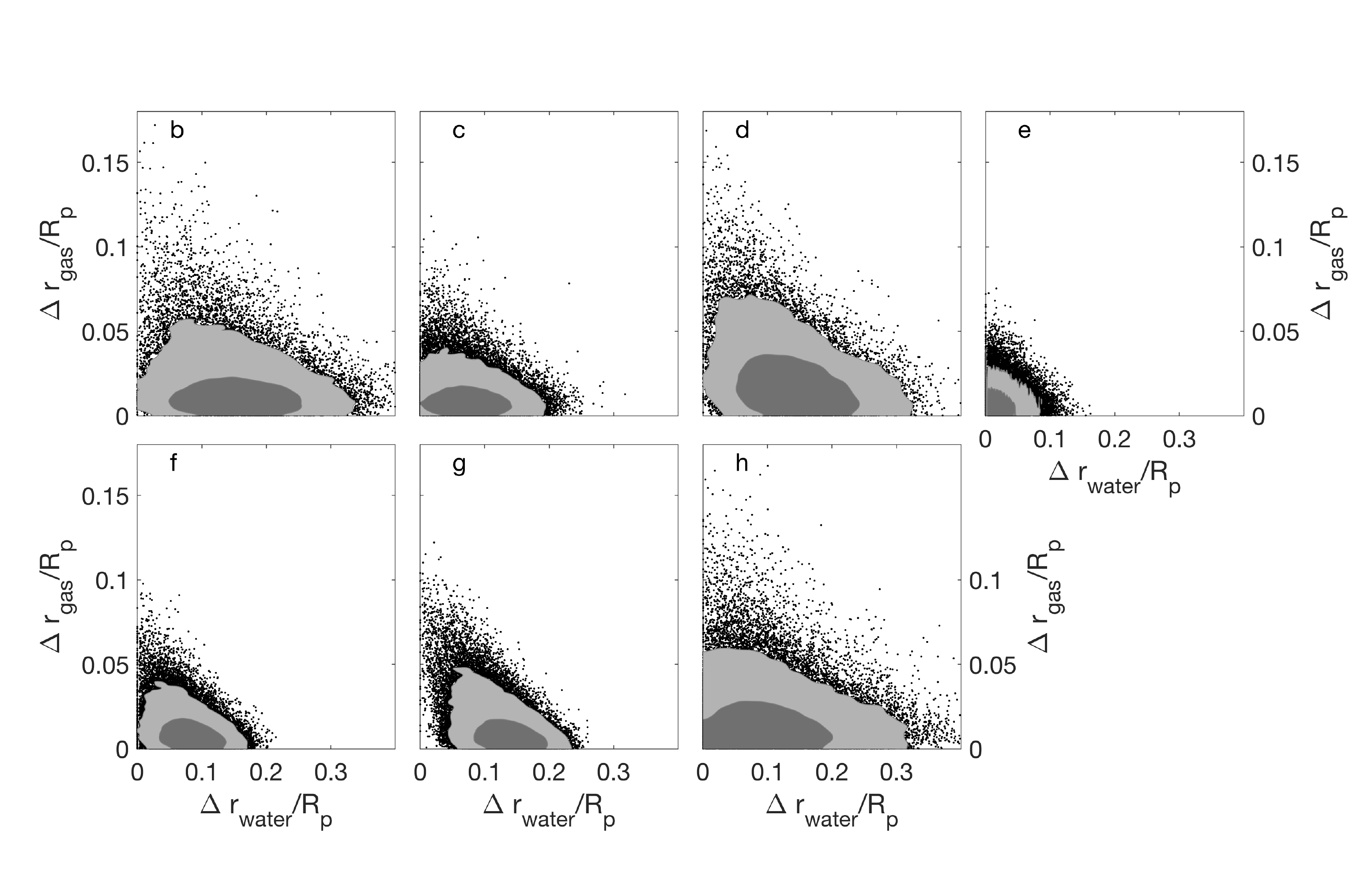}\\
 \caption{Two-dimensional marginalized posterior distribution of gas and water radius fraction for scenario \U and all seven planets (b-h). The difference in radius attributed to the gas or water layers are shown. Dark and light grey regions refer to 1-$\sigma$ and 2-$\sigma$ regions.}
 \label{Fig:uwater}
\end{figure*}

\begin{figure*}[ht]
\centering
\includegraphics[width = .8\textwidth, trim = 0cm 0cm 0cm 0cm, clip]{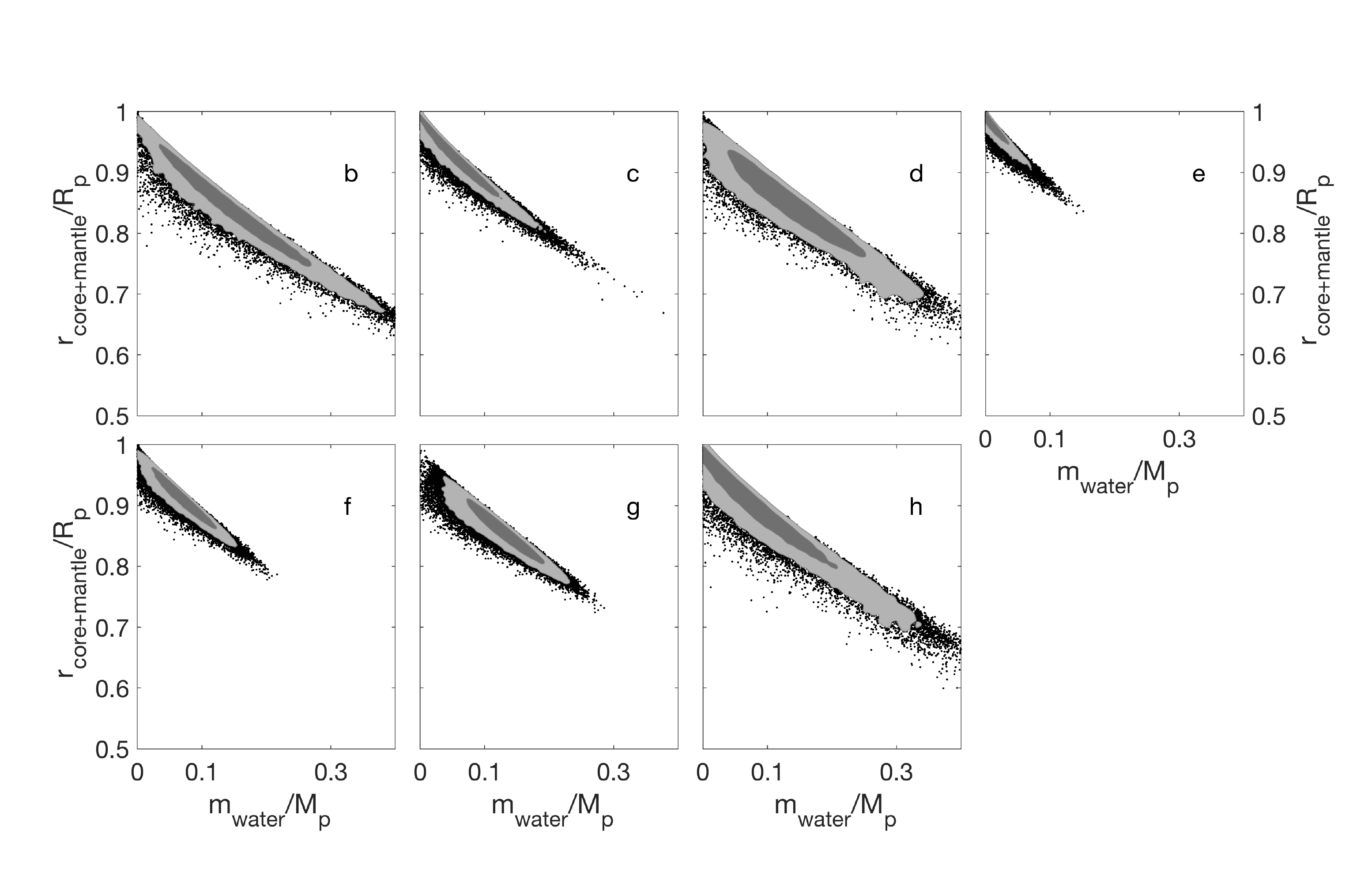}\\
 \caption{Two-dimensional marginalized posterior distribution of rocky interior size and water mass fraction for scenario \U and all seven planets (b-h). Dark and light grey regions refer to 1-$\sigma$ and 2-$\sigma$ regions.}
 \label{Fig:water_rsolid}
\end{figure*}

\subsection{Interior predictions}

Our calculated interior predictions account for the variability in layer thicknesses (of core, mantle, ice/ocean, and gas), layer compositions (of mantle and gas), and thermal states.
In the following, we present one-and two-dimensional marginalized distributions of estimated posteriors. Note, that the 1-$\sigma$ uncertainty regions differ between a distribution that is marginalized over one or two dimensions. The uncertainty regions are generally smaller for one-dimensional marginalized distributions. Stated uncertainty regions refer to one-dimensional marginalized distributions. 

In Figure \ref{Fig:uwater} we show the predicted ranges of radius fractions for gas and water layers for the scenario \U, in which we use the stellar abundance proxy as suggested by \citet{unterborn2017constraining}. 
Possible radius fractions of gas envelopes range between 0-5 \%, while water layers can contribute up to 30\% to the total radius. The uncertainty ranges on possible amounts of water are large. This is because the degeneracy with gas envelope thicknesses (Figure \ref{Fig:uwater}), but mainly the size of the rocky interiors (Figure \ref{Fig:water_rsolid}) is large.
There is a strong correlation between the possible amount of water and the size of the rocky interior, as expected.

In the following, we discuss how much the predicted water mass fractions depend on model assumptions and considered data.


\begin{figure*}[ht]
\centering
\includegraphics[width = .8\textwidth, trim = 0cm 0cm 0cm 0cm, clip]{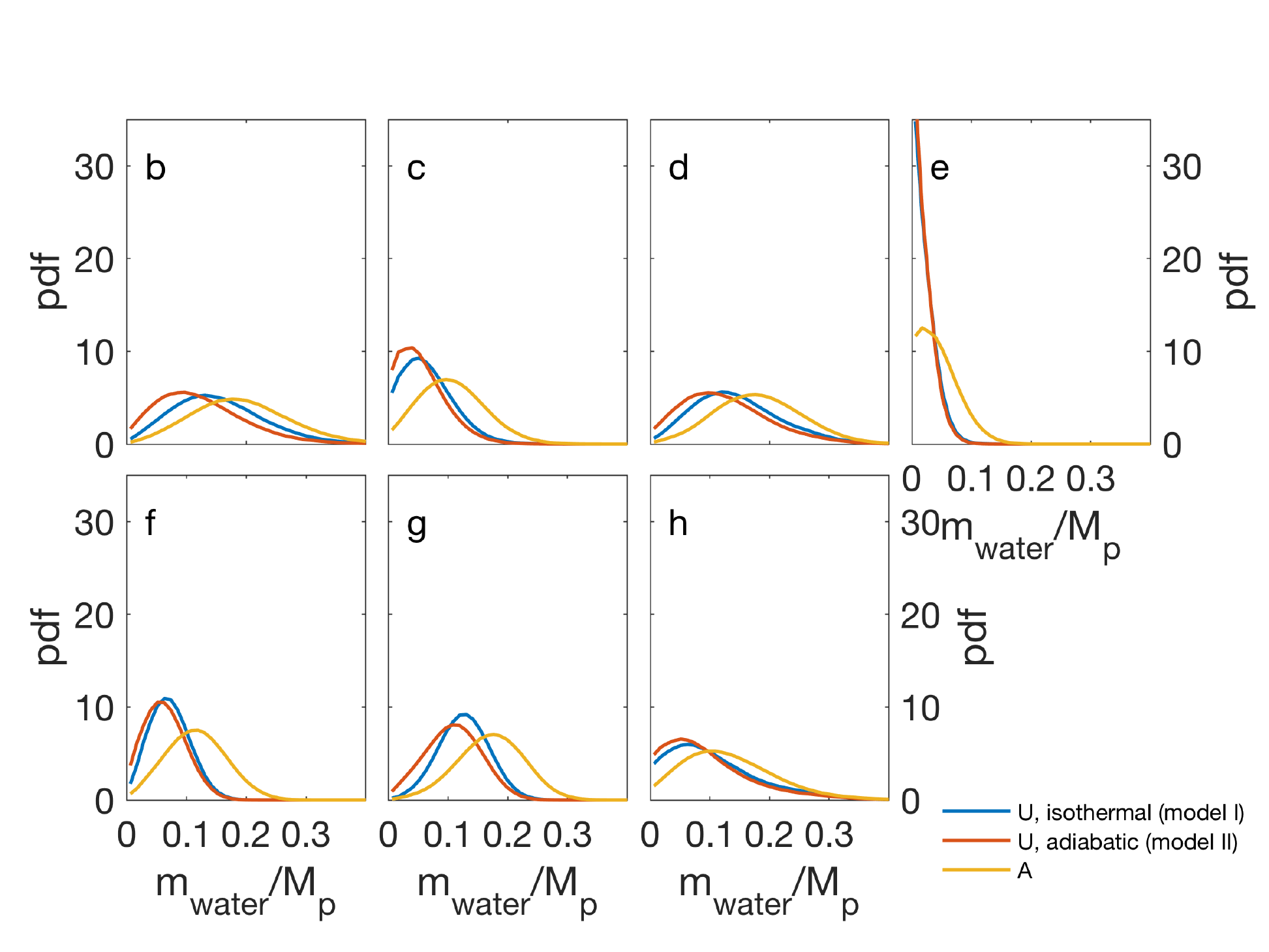}\\
 \caption{One-dimensional marginalized posterior distribution of water mass fractions for all seven planets (b-h). The shown data scenarios use the stellar abundance proxy (\U) or the abundance proxy based on TRAPPIST-1~e (\A). For the gas layer, model I (blue and yellow curve) assumes an isothermal temperature profile, while model II (red curve) assumes a convective adiabat in the gas layer.  }
 \label{Fig:Fwater}
\end{figure*}

\subsubsection{Influence of the choice of abundance proxy}
\label{proxi}
Figure \ref{Fig:Fwater} illustrates how the choice of abundance constraints influences the predicted amount of water. If the stellar proxy (\U) is used, rocky interiors are less dense compared to the proxy that is based on TRAPPIST-1~e (\A). Consequently, the predicted  amount of water is smaller in scenario \U, since less water can be added on top of low density rocky interiors while still fitting mass and radius. There are differences in the median predicted water mass fraction that range from 25\% up to 50\% between \U and \A. For all but planet e, the differences are largely around 30\%. For planet e, it is the discrepancy between mass and radius versus stellar proxy that leads to such large changes in $m_{\rm water}/M_p$ between the scenarios \U and \A.
 Table \ref{resulttable} summarizes the estimated interior parameters and their 1-$\sigma$ uncertainties.

\begin{figure*}[ht]
\centering
\includegraphics[width = .8\textwidth, trim = 0cm 0cm 0cm 0cm, clip]{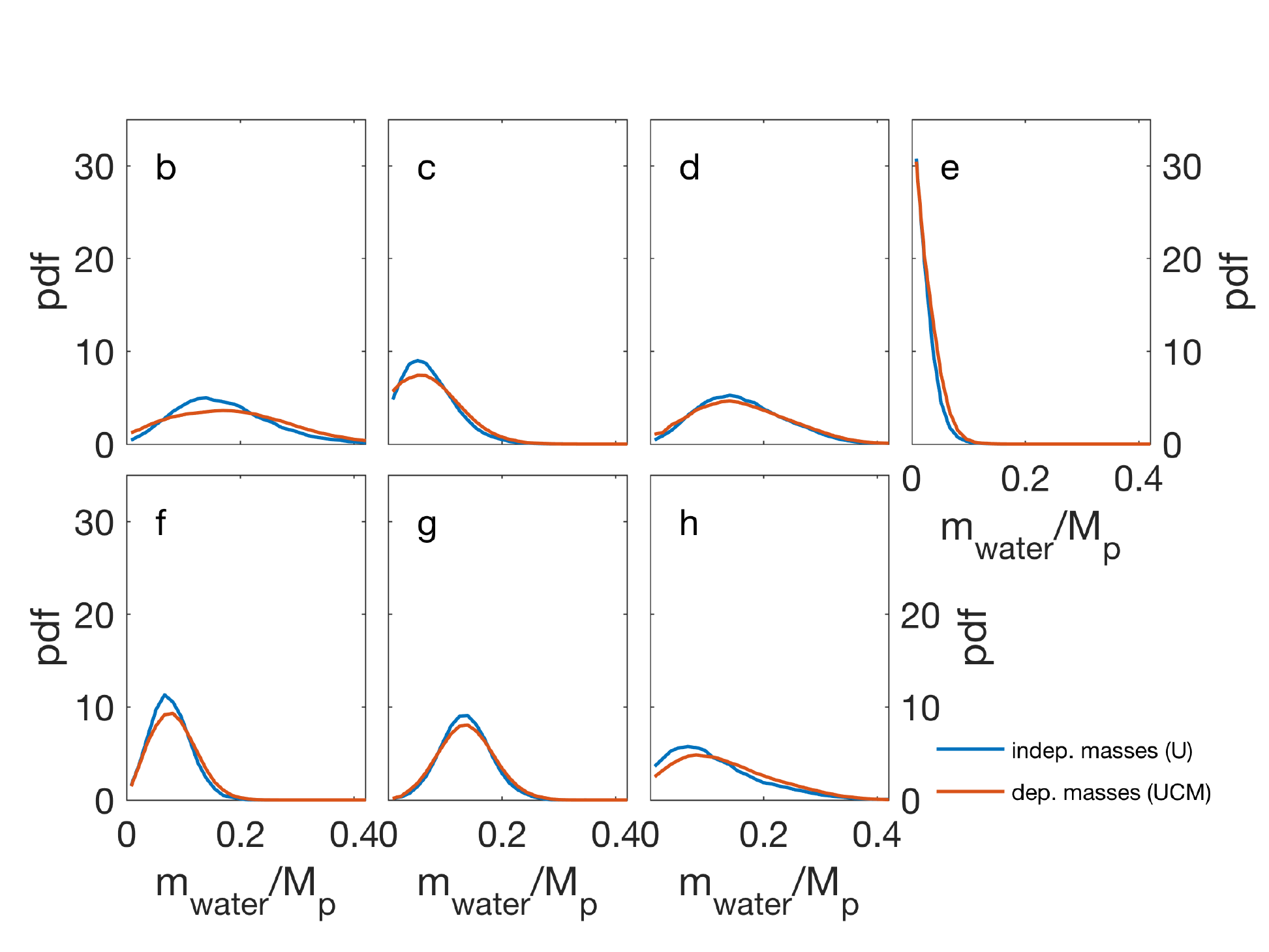}\\
 \caption{One-dimensional marginalized posterior distribution of water mass fractions for all seven planets (b-h). \U ignores interdependencies of planetary data, while \UCM accounts for interdependent masses. The change between \U and \UCM depend on the abundance proxy used. Shown scenarios use the stellar proxy. }
 \label{Fig:Fwater_UC}
\end{figure*}

\begin{figure}[ht]
\centering
\includegraphics[width = .5\textwidth, trim = 1cm 0cm 1.6cm 0cm, clip]{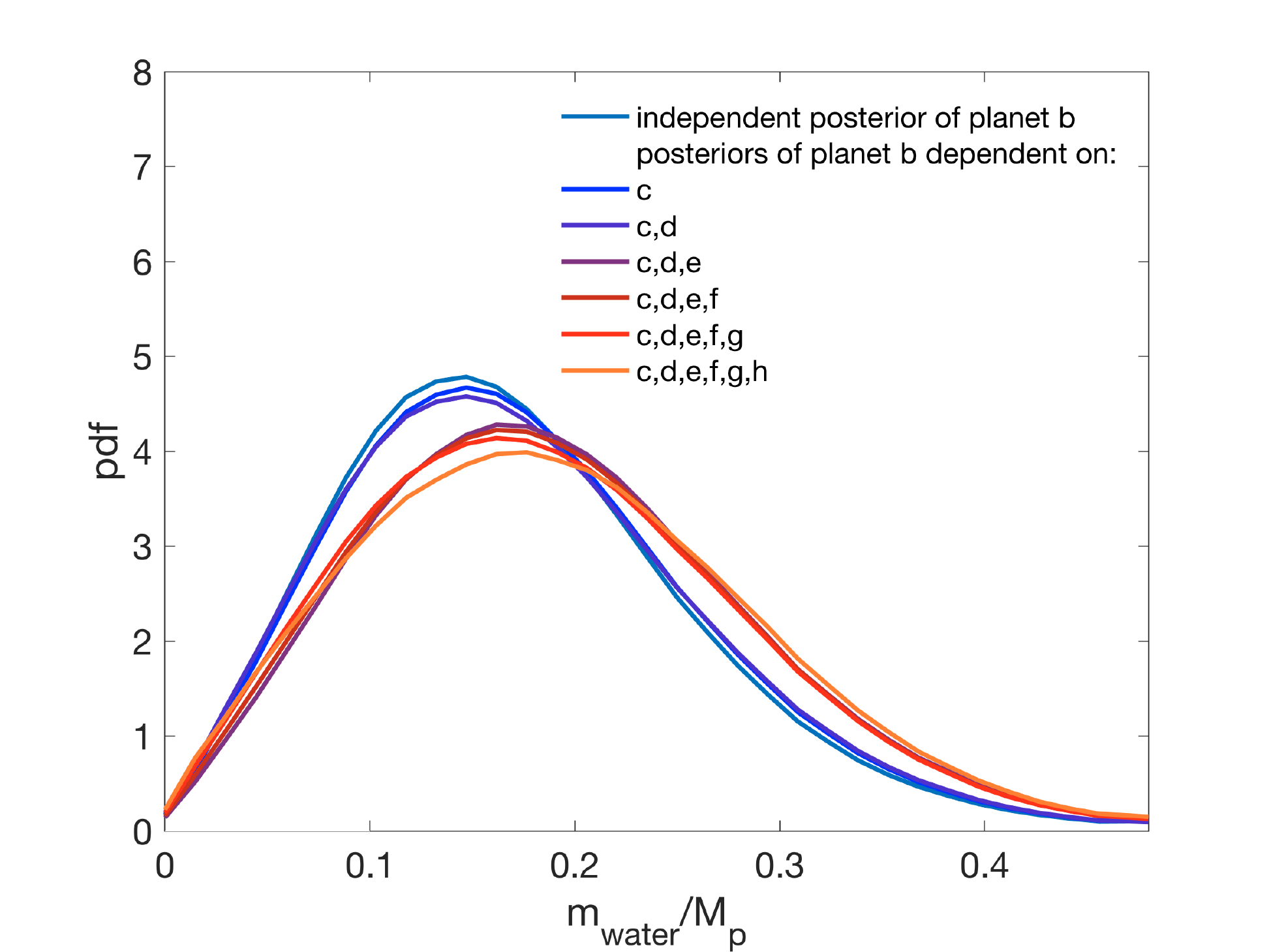}\\
 \caption{One-dimensional marginalized posterior distribution of water mass fractions for planet b. The independent posterior of scenario \U (grey-blue line) yields highest water mass estimates. While subsequently taking into account the interdependency of planetary masses between planet b and the other planets, the curves are colored with more reddish colors and move to lower water mass fractions. The lowest water mass fractions represent scenario \UCM.}
 \label{Figtest}
\end{figure}

\begin{figure}[ht]
\centering
\includegraphics[width = .45\textwidth, trim = 0cm 0cm 1.cm 0cm, clip]{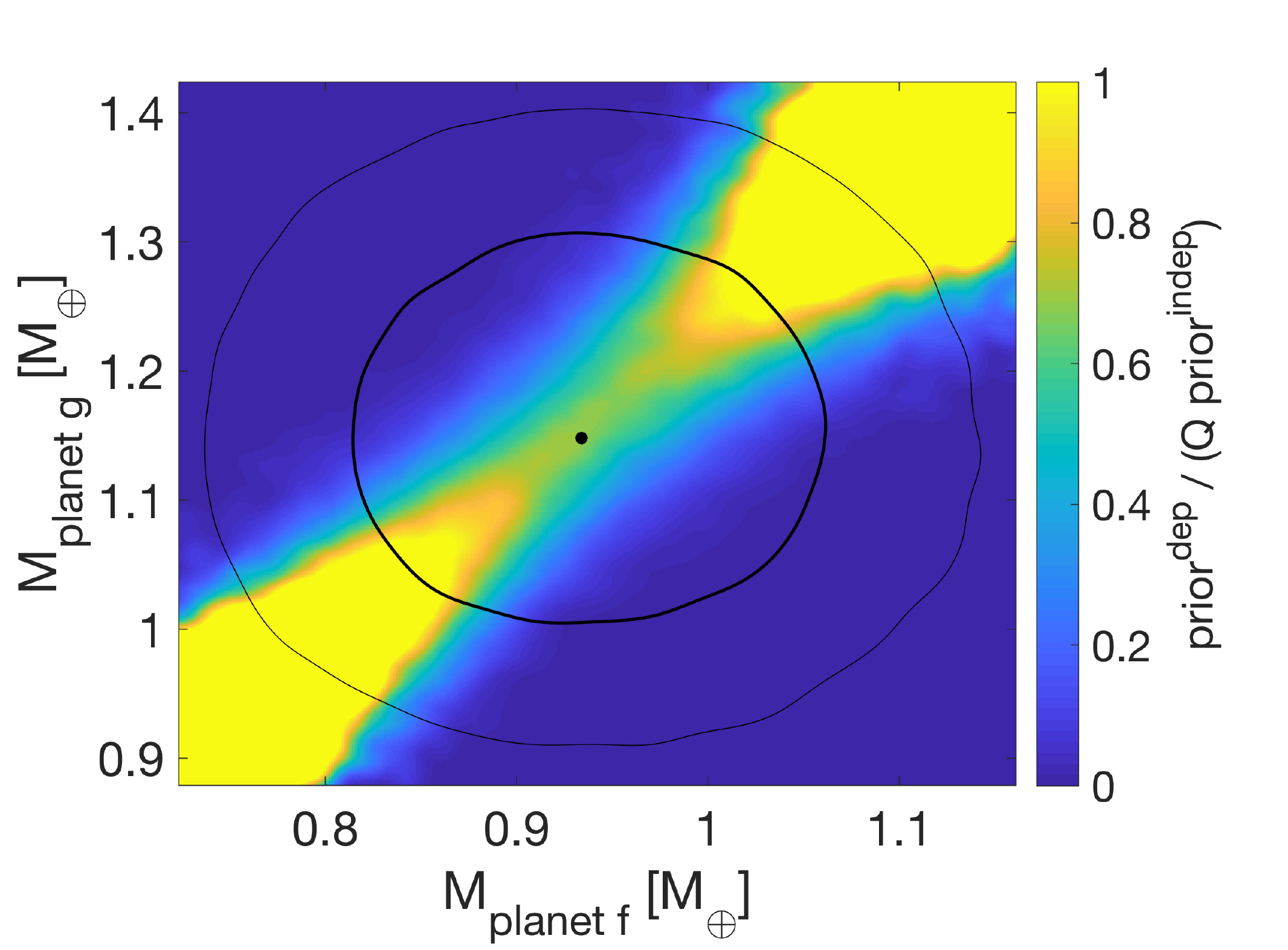}\\
\includegraphics[width = .45\textwidth, trim = 0cm 0cm 1.cm 0cm, clip]{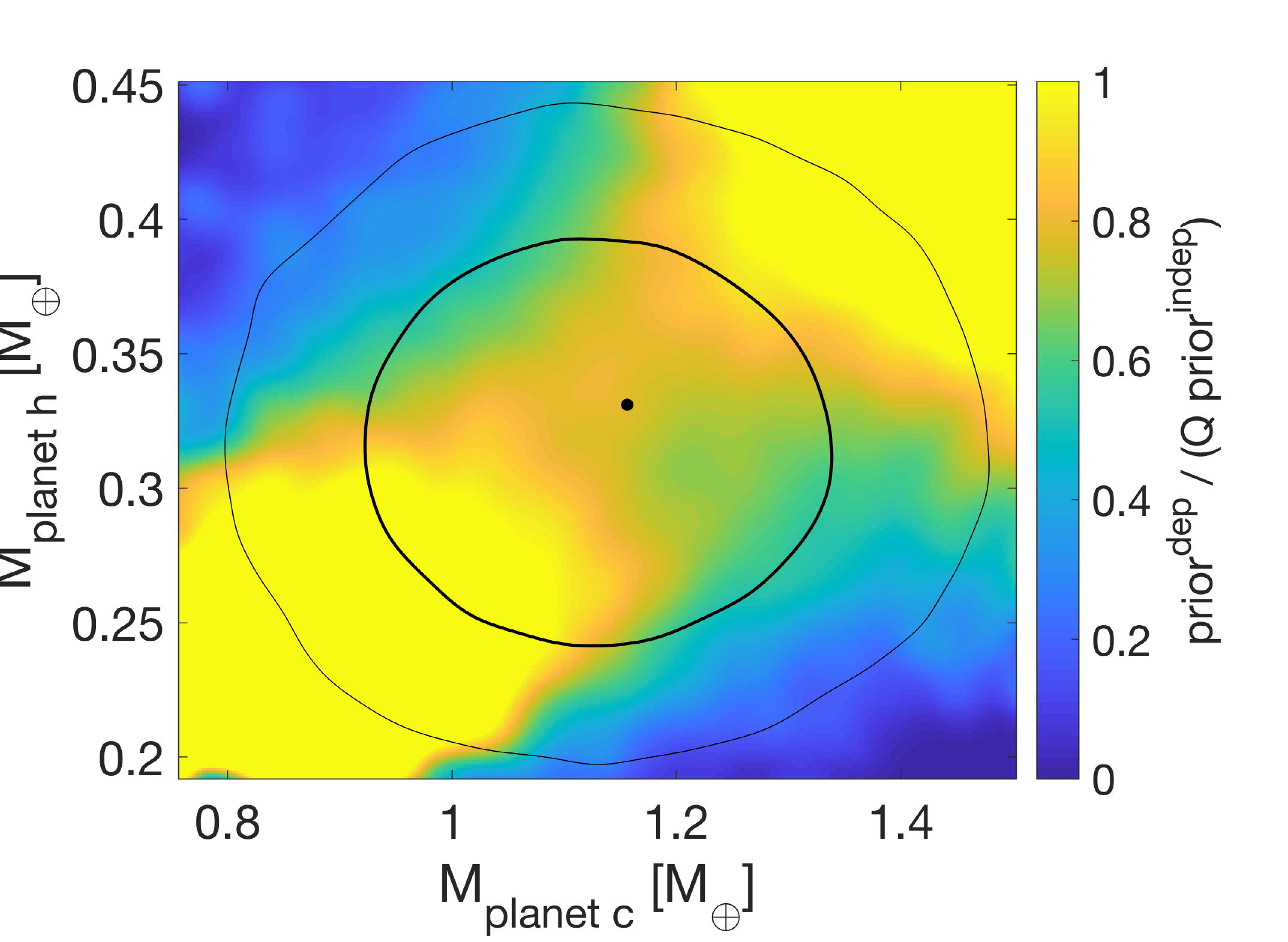}\\
 \caption{\rev{Planetary mass distributions for (upper panel) planets f and g, and (lower panel) planets c and h. The 1- and 2-$\sigma$ contours in white are the independent posteriors of scenario \U. These posteriors are resampled according to the ratio prior$^{\rm dep}/ (Q {\rm prior}^{\rm indep})$ (in color) as described in Section \ref{sec:resamp}  in order to obtain the dependent posteriors.}}
 \label{MRexplain}
\end{figure}

\subsubsection{Accounting for the interdependency of planetary masses}
\label{interdep}

\rev{Inferred planetary masses by \citet{Grimm2018} are significantly correlated between different planets (Figure \ref{fig:CM}).} We account for this interdependency of planetary masses by resampling the posterior interior samples of scenario \U with our new resampling scheme (Section \ref{sec:resamp}) that yields the posterior distribution \UCM. The difference between the posterior distribution \U and \UCM demonstrates the information contained in the interdependency of planetary masses. 
Figure \ref{Fig:Fwater_UC} shows that the differences are \rev{largest for planets b, and h with up to 20 \% difference in both the median water mass fraction and the corresponding uncertainty} (Table \ref{resulttable}).

The information contained in the interdependency of planetary data depends on the specific correlation between the planets. In Figure \ref{Figtest} we show \rev{how the distribution of water mass fraction for planet b changes} while adding step-by-step the information on the correlation between the planetary masses. \rev{For planet b, the information gained by accounting for interdependency of masses is mainly kept in the inter-dependency with the planets e and h. In general, how much information can be gained, depends on the actual level of interdependency of planet pairs. While accounting for inter-dependent data, water mass fractions are shifting towards lower values and uncertainties increase.  }

\rev{Why do uncertainties on estimated water mass fractions increase? The reason is illustrated in Figure \ref{MRexplain} showing  the independent posteriors (\U) and the prior ratio (${ \rho_{d_{ij}}(\dat_i,\dat_j) }/{ Q \rho_{d_i}(\dat_i)\rho_{d_j}(\dat_j)}$ from Equation \ref{eq_res}) that is used to resample from the independent posteriors in order to obtain dependent posteriors. The ratio is highest at the two tailes of the 2D-distributions, where planetary masses are both small or large. These tailes of the independent mass distribution are preferentially resampled. For a given radius, high masses are characterized by low water contents and vice versa. In consequence, the distribution of possible water mass fractions broadens.}

\rev{Why higher water mass fractions are preferentially resampled?} For some of the planets, the highest possible masses cannot be well-described by the stellar proxy (\U). This is the case for planet e (as discussed earlier for Figure \ref{Fig:T1e}) but also c and h (Figure \ref{MRexplain}, lower panel). It is not the case for example for planets f and g (Figure \ref{MRexplain}, lower panel). The consequence is clear from the posteriors (\U) that are shifted to smaller planetary masses compared to the prior data (for e, h, and c). During the resampling of the posterior \U, only those samples survive for which the prior ratio is large. \rev{This is the case for lower masses  (Figure \ref{MRexplain}, lower panel) that are characterized by higher water contents (for a given radius). }

The difference between \U and \UCM is due to the partial incompatibility of planetary data (mass and radius versus the stellar proxy) of e, h, and c, but also the fact that the 2-D mass distributions (Figure \ref{fig:CM}) are not perfect ellipses.


\begin{figure}[ht]
\centering
\includegraphics[width = .5\textwidth, trim = 0cm 0cm 0cm 0cm, clip]{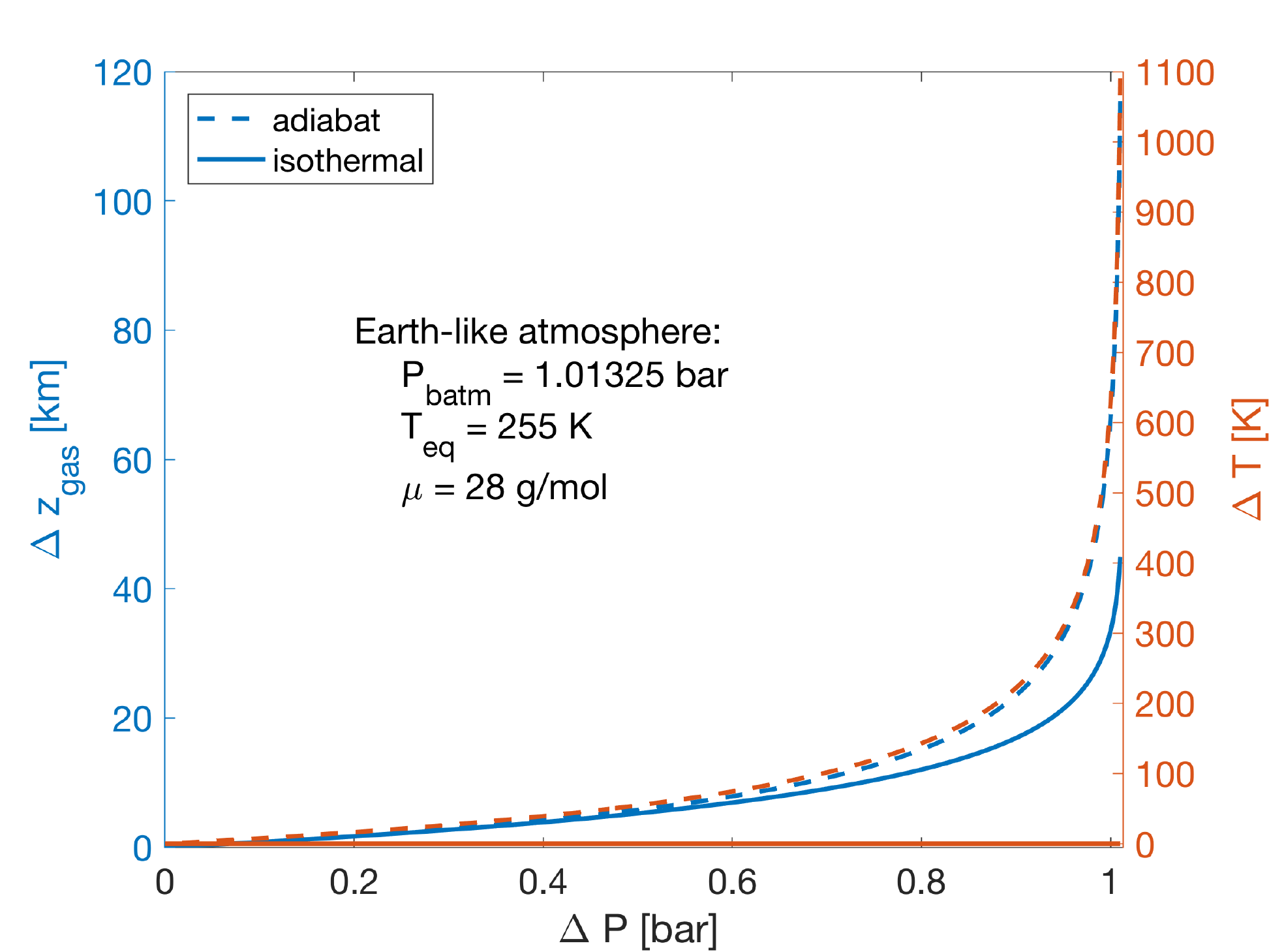}\\
 \caption{Comparison of atmospheric models I (isothermal temperature profile, solid lines) and II (adiabatic, dashed lines) for an Earth-like atmosphere. $\Delta z_{\rm gas}$ is the depth into the gas envelope reaching different pressures $\Delta P$  compared to 20 mbar at $z_{\rm gas} = 0$km. Depending on the model (I or II), the increase in temperature $\Delta T$ is zero or higher than Earth's atmospheric temperatures. The models enclose the average Earth's atmosphere profile and represent end-member cases.}
 \label{Fig:atmos}
\end{figure}

\subsubsection{Influence of atmospheric model}

Figure \ref{Fig:Fwater} shows that the differences in water mass fractions are small when we change from an isothermal to a fully adiabatic temperature model for the gas layer. In the isothermal model I, we have introduced a parameter $\alpha$,  that accounts for cooling and heating of an atmosphere. However, temperatures are largely limited to 400 K and thus, especially for the innermost planets, this model I is not able to produce temperatures as high as suggested by \citet{Grimm2018} (surface temperatures of 750-1500 K for planet b). Model II overcomes this limitation but overpredicts the temperatures in the gas layer by assuming a fully convective gas layer. In consequence, surface temperatures in model II can be very high and can range for example on planet b from 500 - 10.000 K for surface pressures of $1 - 10^{4}$ bars. Thus, while temperatures in model I are underestimated, they are overestimated in model II. This is illustrated in Figure \ref{Fig:atmos} showing pressure-temperature profiles for an Earth-like atmosphere. Global climate models that solve for radiative transfer as presented in \citep{turbet2017climate} yield much more realistic surface temperatures. However, we find that there is only limited influence due to the choice of gas layer model on the estimated mass of the underlying water layer. Median estimates generally vary by 10-15\%, expect for planets b and c with 25\%. The differences are relatively small compared to other influences (Section \ref{proxi} and \ref{systematics}).

Estimated water mass fractions are smaller for model II compared to model I. This is because the lower gas envelope temperatures in model I results in smaller scale heights, thin gas envelopes and thus larger amounts are required to fit planet mass and radius. The solutions given the two models I \& II represent end-member solutions. 


\begin{figure*}[ht]
\centering
\includegraphics[width = .8\textwidth, trim = 0cm 0cm 0cm 0cm, clip]{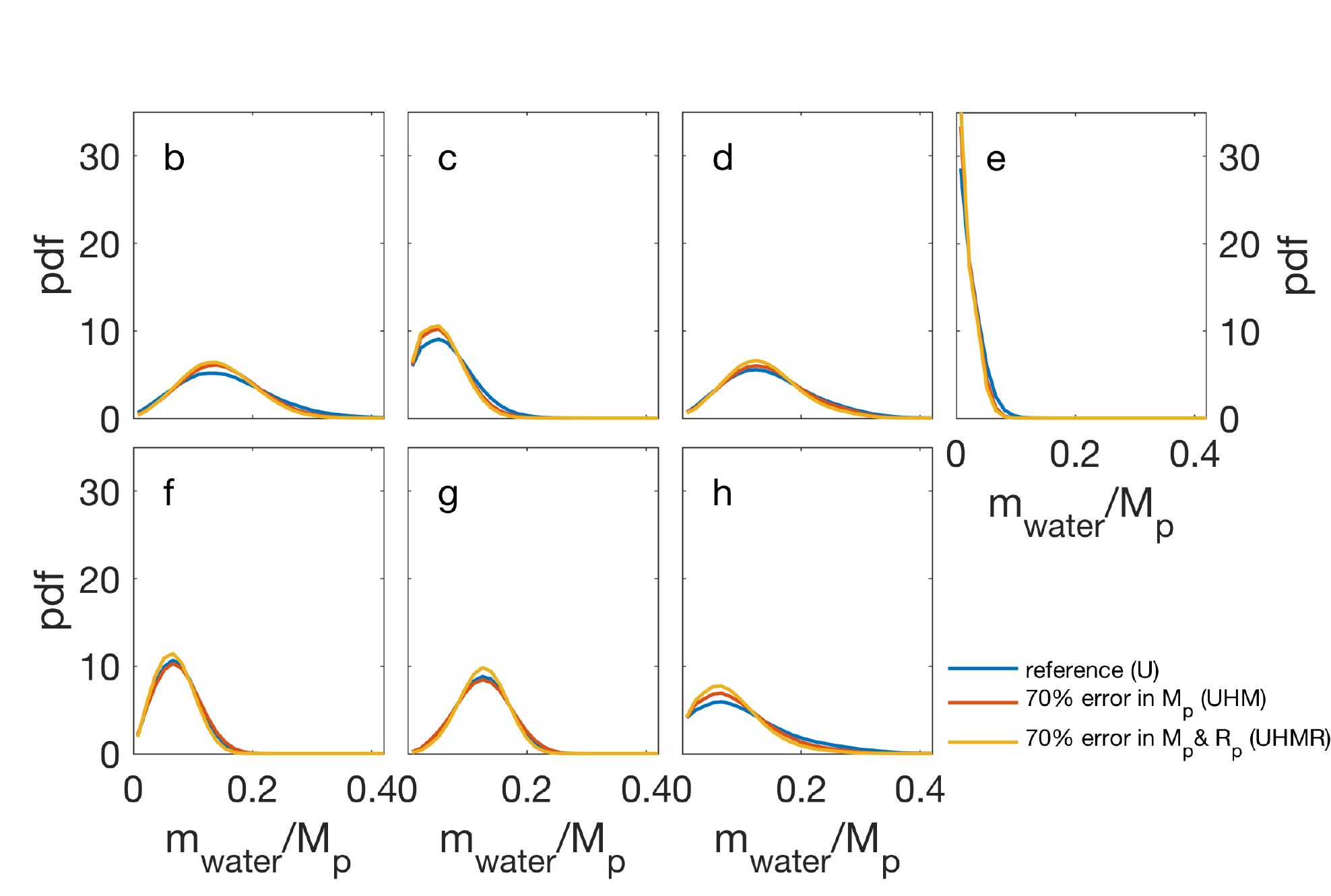}\\
 \caption{One-dimensional marginalized posterior distribution of water mass fractions for all seven planets (b-h).  For \U the nominal data uncertainties are used, while for \UHM and \UHMR 70 \% of the uncertainties on planet mass and on mass \& radius are used, respectively. All shown data scenarios use the stellar proxy.}
 \label{Fig:Fwater_HM}
\end{figure*}

\subsubsection{Relevance of data uncertainty}
\label{sec:data}
Figure \ref{Fig:Fwater_HM} illustrates how much our ability of constraining water mass fraction would improve by more precise estimates of mass and radius. An improved estimate on planet mass (70\% of the nominal uncertainty) only has generally marginal influence on the median estimate of $m_{\rm water}/M_p$ but it reduces the 1-$\sigma$ confidence interval by 5-20\%. Improved estimates on both planet mass and radius (70\% of the nominal uncertainty) reduces the 1-$\sigma$ confidence interval by 10-35\%. Largest improvements are seen for planet e and h, while smallest improvements are seen for f and g.
The information gained by improved data uncertainty is small compared to the different choices of abundance proxies (Figure \ref{Fig:Fwater}) as well as the information kept in the interdependency of planetary data (Figure \ref{Fig:Fwater_UC}).

\begin{table*}[ht] \setlength{\tabcolsep}{8pt}
 \caption{Interior parameter estimates with 1-$\sigma$ uncertainties of the 1-D marginalized posterior distributions. \label{resulttable}}
 \widesplit{%
\begin{tabular}{l|lll |lll| lll| lll |lll| lll| lll| }
\hline\noalign{\smallskip}
planet: \row{& \multicolumn{3}{c}{TRAPPIST-1b}& \multicolumn{3}{c}{TRAPPIST-1c}& \multicolumn{3}{c}{TRAPPIST-1d}}{& \multicolumn{3}{c}{TRAPPIST-1e}& \multicolumn{3}{c}{TRAPPIST-1f}& \multicolumn{3}{c}{TRAPPIST-1g}}{& \multicolumn{3}{c}{TRAPPIST-1h}}\\
scenario: \row{& \U &  \A & \UCM & \U &  \A & \UCM & \U &  \A & \UCM }{& \U &  \A & \UCM & \U &  \A & \UCM & \U &  \A & \UCM }{& \U &  \A & \UCM}\\
\noalign{\smallskip}
\hline\noalign{\smallskip}
\rcore/\rsolid\row{&$0.39_{-0.11}^{+0.09}$&$0.50_{-0.14}^{+0.12}$&$0.39_{-0.11}^{+0.09}$&$0.40_{-0.11}^{+0.08}$&$0.52_{-0.13}^{+0.11}$&$0.40_{-0.11}^{+0.08}$&$0.39_{-0.11}^{+0.09}$&$0.51_{-0.14}^{+0.12}$&$0.39_{-0.12}^{+0.08}$}{&$0.45_{-0.10}^{+0.06}$&$0.58_{-0.11}^{+0.09}$&$0.45_{-0.10}^{+0.07}$&$0.39_{-0.11}^{+0.08}$&$0.52_{-0.13}^{+0.12}$&$0.39_{-0.11}^{+0.08}$&$0.39_{-0.11}^{+0.08}$&$0.50_{-0.14}^{+0.12}$&$0.39_{-0.11}^{+0.09}$}{&$0.40_{-0.11}^{+0.08}$&$0.52_{-0.14}^{+0.11}$&$0.40_{-0.12}^{+0.08}$}\\ 
\rsolid$/R_p$\row{&$0.84_{-0.06}^{+0.06}$&$0.79_{-0.07}^{+0.07}$&$0.82_{-0.07}^{+0.08}$&$0.92_{-0.05}^{+0.04}$&$0.86_{-0.06}^{+0.06}$&$0.91_{-0.05}^{+0.04}$&$0.84_{-0.06}^{+0.05}$&$0.79_{-0.06}^{+0.06}$&$0.84_{-0.06}^{+0.06}$}{&$0.97_{-0.02}^{+0.02}$&$0.93_{-0.05}^{+0.04}$&$0.97_{-0.02}^{+0.02}$&$0.91_{-0.03}^{+0.03}$&$0.86_{-0.05}^{+0.05}$&$0.91_{-0.03}^{+0.03}$&$0.86_{-0.03}^{+0.03}$&$0.81_{-0.05}^{+0.05}$&$0.86_{-0.03}^{+0.03}$}{&$0.89_{-0.07}^{+0.06}$&$0.84_{-0.07}^{+0.07}$&$0.87_{-0.07}^{+0.06}$}\\ 
$m_{water}/M_p$ \row{&$0.15_{-0.07}^{+0.08}$&$0.19_{-0.08}^{+0.09}$&$0.16_{-0.09}^{+0.10}$&$0.06_{-0.04}^{+0.05}$&$0.10_{-0.05}^{+0.06}$&$0.06_{-0.04}^{+0.05}$&$0.14_{-0.06}^{+0.08}$&$0.18_{-0.07}^{+0.07}$&$0.14_{-0.07}^{+0.09}$}{&$0.02_{-0.01}^{+0.02}$&$0.04_{-0.03}^{+0.04}$&$0.02_{-0.01}^{+0.02}$&$0.07_{-0.03}^{+0.03}$&$0.11_{-0.05}^{+0.05}$&$0.07_{-0.03}^{+0.03}$&$0.13_{-0.04}^{+0.04}$&$0.17_{-0.06}^{+0.05}$&$0.13_{-0.04}^{+0.04}$}{&$0.09_{-0.06}^{+0.09}$&$0.12_{-0.07}^{+0.08}$&$0.11_{-0.06}^{+0.09}$}\\ 
$r_{water}/R_p$ \row{&$0.15_{-0.06}^{+0.07}$&$0.19_{-0.07}^{+0.07}$&$0.16_{-0.08}^{+0.07}$&$0.07_{-0.04}^{+0.05}$&$0.12_{-0.06}^{+0.06}$&$0.08_{-0.04}^{+0.05}$&$0.14_{-0.06}^{+0.06}$&$0.18_{-0.06}^{+0.06}$&$0.13_{-0.06}^{+0.07}$}{&$0.02_{-0.01}^{+0.02}$&$0.06_{-0.04}^{+0.05}$&$0.02_{-0.01}^{+0.02}$&$0.08_{-0.03}^{+0.03}$&$0.13_{-0.05}^{+0.05}$&$0.08_{-0.03}^{+0.03}$&$0.13_{-0.04}^{+0.03}$&$0.18_{-0.06}^{+0.05}$&$0.13_{-0.04}^{+0.04}$}{&$0.10_{-0.06}^{+0.08}$&$0.14_{-0.07}^{+0.08}$&$0.11_{-0.06}^{+0.08}$}\\ 
$r_{gas}/Rp$\row{&$0.01_{-0.01}^{+0.01}$&$0.01_{-0.01}^{+0.01}$&$0.01_{-0.01}^{+0.01}$&$0.01_{-0.01}^{+0.01}$&$0.01_{-0.01}^{+0.01}$&$0.01_{-0.01}^{+0.01}$&$0.02_{-0.01}^{+0.02}$&$0.02_{-0.01}^{+0.02}$&$0.02_{-0.01}^{+0.02}$}{&$0.01_{-0.00}^{+0.01}$&$0.01_{-0.01}^{+0.01}$&$0.01_{-0.00}^{+0.01}$&$0.01_{-0.01}^{+0.01}$&$0.01_{-0.01}^{+0.01}$&$0.01_{-0.01}^{+0.01}$&$0.01_{-0.01}^{+0.01}$&$0.01_{-0.01}^{+0.01}$&$0.01_{-0.01}^{+0.01}$}{&$0.01_{-0.01}^{+0.02}$&$0.01_{-0.01}^{+0.02}$&$0.01_{-0.01}^{+0.02}$}\\ 
\end{tabular}%
}
\end{table*}


\begin{figure*}[ht] 
\centering
\includegraphics[width = .85\textwidth, trim = 0cm 0cm 0cm 0cm, clip]{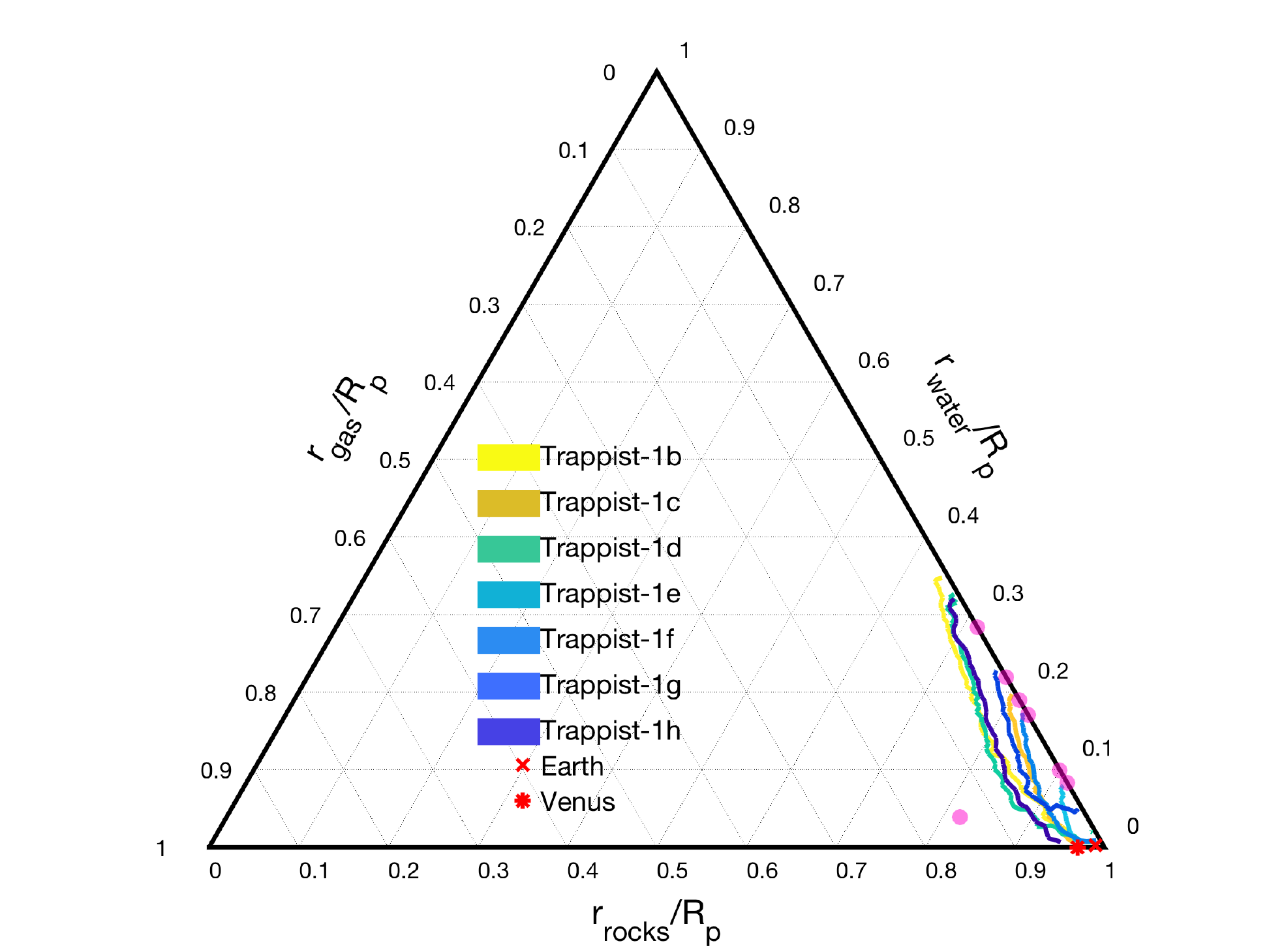}\\
 \caption{Ternary diagram showing the marginalized 2-$\sigma$ distributions of rock-water-gas radius fractions for TRAPPIST-1 planets and scenario \UCM. All three radius fractions sum up to one, which makes it possible to represent the distributions in a ternary diagram. For comparison, Earth and Venus are shown in red, mean estimates for Europa, Ganymede, Callisto, Titan, Triton, Pluto, and Charon are shown in purple. }
 \label{Fig:Triangular plot}
\end{figure*}

\begin{figure}[ht]
\centering
\includegraphics[width = .5\textwidth, trim = 0cm 0cm 0cm 0cm, clip]{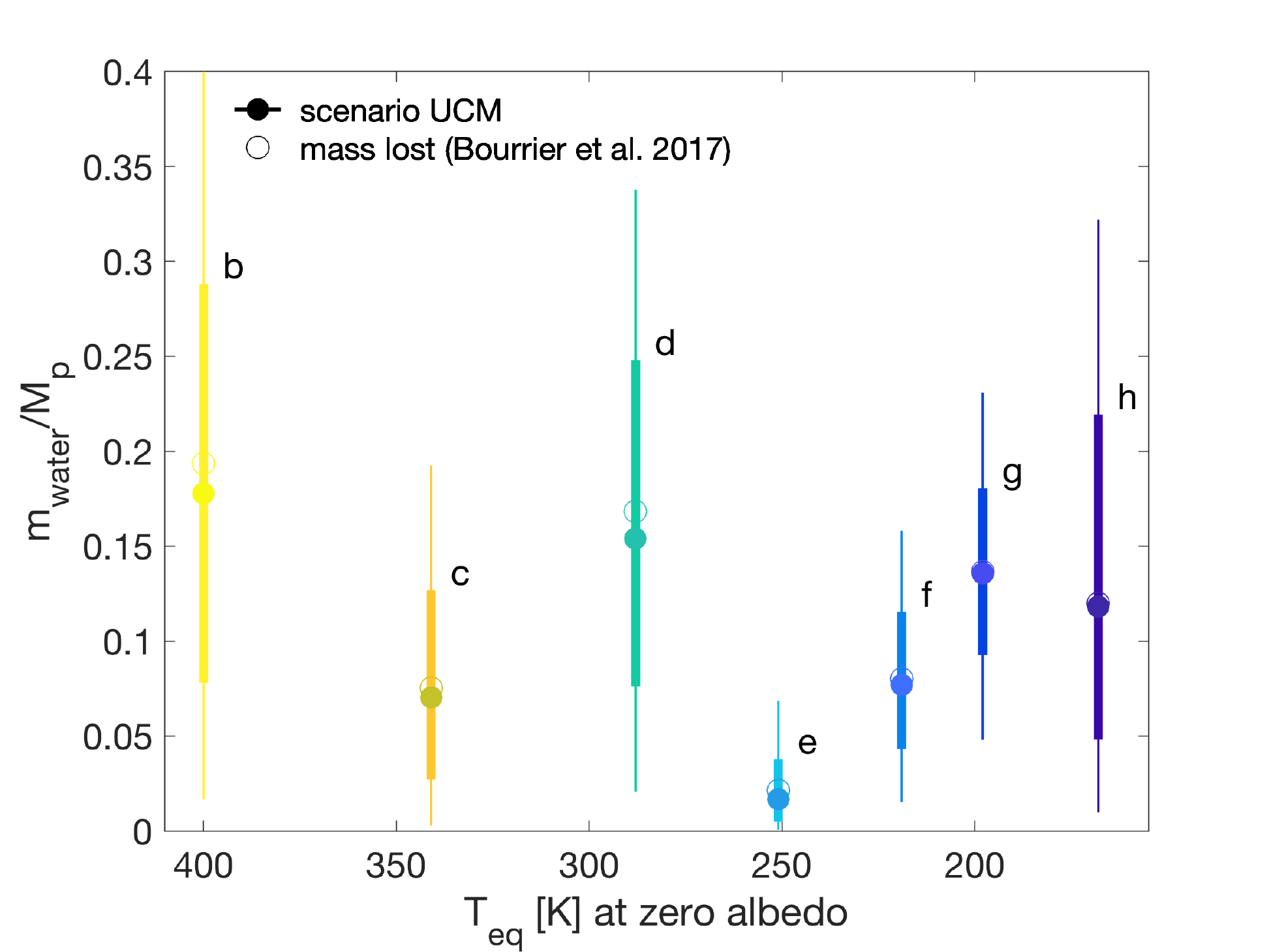}
\includegraphics[width = .5\textwidth, trim = 0cm 0cm 0cm 0cm, clip]{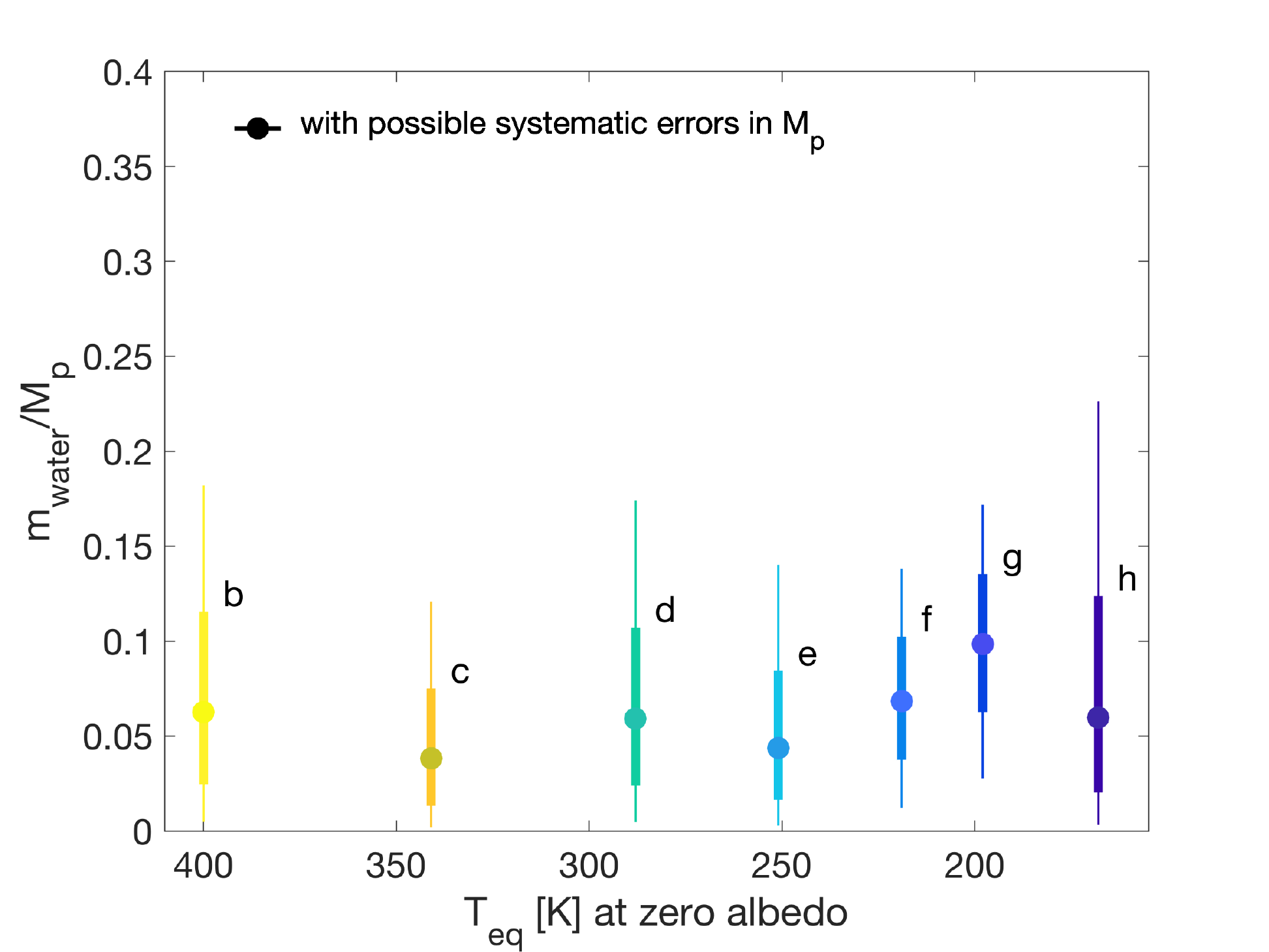}
 \caption{Marginalized water mass fractions as a function of equilibrium temperature T$_{\rm eq}$ (at zero albedo). Upper panel shows scenario \UCM. No obvious trend of increasing water mass fraction with larger orbital distance and thus cooler T$_{\rm eq}$. For the water mass fractions, the 3th-16th-84th-97th-percentiles are depicted by the thin and thick error bars. How much water mass fraction there could have been after formation (open circles) is calculated by adding the amount of lost water \citep{Bourrier_2017} to our inferred median estimates (filled circles).  \rev{Lower panel shows that possible systematic shifts in planet masses within $\pm$ 24\% could in fact result in finding water mass fractions that are uniform or increasing with orbital distance.}}
 \label{Fig:simple}
\end{figure}

\section{Discussion}
\label{discussion}

Although masses, radii, and temperature conditions of the TRAPPIST-1 planets generally remind us of terrestrial Solar System planets, the possible interiors of the TRAPPIST-1 planets can contain significant water budgets of up to 20 or even 30 \%, unlike the terrestrial planets ($\le$0.02 \% surface water). This is because of the roughly 20\% lower bulk densities, their associated uncertainties, and the large degeneracy due to variability in structure and composition of the rocky interior, as well as in gas layer characteristics.

 \subsection{Systematic data biases}
\label{systematics}
\rev{Recent announcements \citep{Demoryconf} have indicated that previously unavailable observations for the TTV analysis of \citet{Grimm2018} provide a higher accuracy with changes in planetary masses within $\pm$ 24\%. In Figure \ref{Fig:simple}, we show that consequences for the estimated water mass fractions are large. Planet e would not be the odd case of a super-Mercury. All planets could be consistent with an increasing water mass fraction with orbital period (within 2-12 \% with respect to 1-$\sigma$ errors) or a uniform water mass fraction of ~7\% (within 1-$\sigma$ errors). Such large amounts of water would exclude Earth-like habitability even for the temperate planets since their rocky interiors are separated by icy layers from the water oceans in that case. In contrast, the published data estimates \citep{Grimm2018}, suggests no clear trend of water mass fraction with orbital distance as discussed below (Figure \ref{Fig:simple}). 

Specifically for the TRAPPIST-1 planets, the trend of changing water mass fractions with systematic shifts in planetary masses can be described as follows. A shift of +25\% in planetary mass can reduce the water mass fraction by 65 \% (i.e., the reference water mass fraction multiplied by a factor of 0.35). For shifts of +20\%, +10\%, +5\%, and -15\% the factor is roughly 0.4, 0.55, 0.75, and 2.6 respectively. 

In general, biases from the TTV analysis are extremely difficult to quantify because they stem from limited observations.
Biases from TTV can be caused by unresolved degeneracies between eccentricity, mass and arguments of perihelion of the different planets. Limited observation time can cause the TTV analysis to prefer solutions which depart from the true solution, or the parameter search is trapped in a local minimum of the sampling parameter space.  Also strong tidal effects, nearby perturbing stars or even yet undetected additional planets can introduce a TTV signal bias.  Further TTV analysis on a larger amount of transit observations are needed to increase the accuracy of planetary masses.}

 \subsection{Rocky interiors and TRAPPIST-1~e}
The degeneracies within structure and composition of rocky interiors stem from the fact that we allow for variable silicate mantle compositions and core sizes. Independent constraints on element abundances as taken from stellar proxies are highly valuable for refined rocky interior estimates  \citep{dorn2015can}. Here, however, no direct estimates from the faint host star are available. Instead we have adopted two different abundance proxies (see Table \ref{tabledata}): a stellar proxy (scenario \U) as derived from F-G-K stars of similar metallicities  \citep{unterborn2017constraining} and a proxy based on only mass and radius of the most dense planet of the system: planet e  (scenario \A).

Depending on the abundance proxies used for refractory elements (\U or \A), the interior estimates change with significant influences on  the predicted possible water budgets (Section \ref{proxi}). Any decision on whether scenario \U or \A is more likely, depends on the interpretation of planet e.

If TRAPPIST-1~e was a super-Mercury type planet, it would not represent the rocky analogue for all planets. In this case, scenario \A would be best to only describe planet e, while all other planets would be best described by scenario \UCM~\rev{(in case the stellar proxy would be excluded for planet e)}. A super-Mercury type interior could be explained by a high-energy giant impact \citep{benz1988,stewart2009velocity,marcus2010minimum}. This would also allow to explain why inferred interiors for planet e are poorest in volatiles, although its neighboring planets are volatile-rich. A mantle-stripping impact would require large impact velocities or small impact angles \citep{stewart2009velocity}, or an impactor larger than planet e, which does not exist in the system.  It remains to be studied if impacts are a reasonable scenario. 

If TRAPPIST-1~e was not a super-Mercury type planet, there are two interpretations possible. First, TRAPPIST-1~e can be indeed a rocky analogue of all the planets regarding the relative ratios of rock-forming elements. In this case, all planets have compositions that are not Earth-like implying that the iron was more abundant compared to solar compositions. In this case, all planetary interiors are best described by scenario \A.

Second, let us assume that TRAPPIST-1~e is indeed similar to an Earth interior and thus well represented by \U, but its mass and radius estimates are effected by systematic biases (as discussed in Section \ref{systematics}). 
In this case, scenario \U may well describe planet e, while dismissing high bulk-density interiors. Thus many \rev{mass-radius-realizations} are incompatible for planet e in scenario \U. Given that all planetary masses (radii)  are correlated to the mass (radius) of planet e, there are also many planetary masses (radii) of the other planets that are incompatible with \U. In consequence, accounting for the interdependencies of planetary data improves interior estimates. If this case is true, all interiors are well described by scenario \UCM.  \\

 \subsection{Water budgets}

The layers of water can generally be in liquid, solid, and even super-critical state. Water vapor is excluded as we impose that any water layer requires a gaseous layer on top that imposes a pressure at least as high as the vapor pressure of water. For simplification, we use pure water and neglect other forms of ices (e.g., CO$_2$).
For all but planet e, most of the possible planetary interiors are characterized by ice layers separating water oceans from rocky interiors. This is due to the fact that either the large amounts of water allow to reach pressures high enough to form high-pressure ices or temperatures are sufficiently cool (especially for the outer planets f, g, h). We note that the thermal states of the planets are unknown and significant tidal heating may allow for more extended thicknesses of liquid water. Planet e is the volatile-poor exception  among the planets.  It's possible tiny amounts of water and temperate conditions allow for liquid water on the surface. 
Also for planets c and h, the uncertainty on mass and radius allows for negibly thin volatile layers.

In Figure \ref{Fig:Triangular plot}, we show the 2-$\sigma$ uncertainty distribution for scenario \U and all planets in comparison with mean estimates for Earth, Venus, but also Titan and icy satellites, that probably formed outside the water ice-line: Europa, Ganymede, Callisto, Triton, Pluto, and Charon. The  water mass fractions of the shown  Solar System objects range from 0 to 30\%  \citep[for a review see][]{nimmo2016ocean} and are comparable to our inferred ranges of possible water mass fractions. Water mass fractions of Solar System objects on the order of 1000 km or smaller can have higher water budgets (Saturn's moon Tethys with a radius of 1060 km is almost entirely composed of water with a bulk density of 0.98 g/cm$^3$.)
Formation models that simulate the structure and evolution of disks  \citep{alibert2016formation} predicted planets to be more volatile-rich around low-mass stars, since the snow-line is located closer to the stars compared to the Solar System. Indeed, the TRAPPIST-1 system is significantly more volatile-rich than the terrestrial Solar-System planets. 

Our results suggest that an upper water mass fraction to be not much larger than 30\%. The amount of water loss \citep{Bourrier_2017} or water delivery \citep{kral2018cometary} during the long-term evolution of the planets after formation is small compared to our predicted uncertainties on water budgets. While water budgets can be altered by 1 to 100 Earth oceans, the predicted water budgets range from some tens to some thousands of Earth oceans with uncertainties on the scale of hundreds to thousands of Earth oceans. Therefore, our inferred water budgets largely represent the water budgets originally accreted during formation.

In comparison to expected water mass fraction of material outside the water snow-line of 50\% \citep{lodders2003solar}, our maximum predicted water budgets are significantly smaller (30\%). This can be due to a mixed accretion of volatile-poor and volatile-rich planetesimals from within and beyond the ice-line, respectively. Sufficient mixing of volatile-poor and volatile-rich planetesimals for all planets is also reasonable given that no clear trend of water budget with orbital distance is seen as shown in Figure \ref{Fig:simple}.
Here, we plot the inferred water mass fractions against the equilibrium temperature at zero albedo, which relates to orbital distance. The error bars on water mass fraction represent 3th-16th-84th-97th-percentiles of the cumulative distribution. Neither a clear trend of water mass fraction with orbital distance nor with planet mass could be identified.  Thus our findings suggest that the accreted material stems from regions in the disk where mixing  was efficient enough to blur any possible trends of increasing water mass fraction with orbital distance, unless migration reordered the planets. Below, we discuss the implications for the formation of the planets in more detail.

 \subsection{Implications for planet formation}

The two main conclusions of our calculations are that the water content is very diverse among the planets, with no obvious correlation with neither the mass or the period, and that the maximum amount of water is somewhat lower than the icy content of planetesimals usually assumed in the Solar System.

The scatter in planetary water budgets  is at odds with what we observe in the Solar System, both in the planets and in galilean satellites. In both cases, innermost objects tend to be dryer than outermost ones. While an uniform water budget in all planets cannot be excluded at the 2-$\sigma$ level (see Fig. \ref{Fig:simple}), this scatter is puzzling. Different origins can be invoked for this scatter. First planets likely migrated during their formation \citep{alibert2016formation}, and therefore accreted planetesimals from different parts in the disk, some dry (inside the iceline) some other wet (beyond the iceline). In addition, the iceline itself is likely to move during the formation of planets, and this can modify the water content of solids in the disk (pebbles but also small planetesimal). In both cases, one can therefore expect that the average water content of accreted solids (which does not necessary reflect the final water content of planets as discussed below) can vary depending on the exact formation path of the planets. What is however not clear, is whether it is possible to end up with a water budget that shows no trend. Indeed, one would expect, from a smooth and ordered migration of planets, that the water
 content should monotonically increase as a function of period, as we see, for example, in the case of Galilean satellites. Breaking such a monotonic trend requires exchanging the place of some planets, which requires a strong level of dynamical rearragement resulting from gravitational interactions. Wether such a re-arragment is possible, while at the same time ending with a flat and resonant system, remains to be investigated.
 
Another possibility is that the final water budget of planets is not the one of the accreted solids. Indeed, during their formation, planets accrete a gas envelope (which maybe lost afterwards). The interaction of the gas envelope, the protoplanetary disk, and the accreted solids can lead, under some circumstances, to the loss of some of the accreted solid material \citep{alibert2017maximum}, however see \citet{brouwers2018cores}. This process results from the advection of gas from the disk to the planetary atmosphere, and such a process could prevent accretion of water by forming planets. As the efficiency of advection depends on different parameters (including planetary mass, gas density in the disk, distance to the star) in a non-trivial way, it is not clear if such a process can destroy a pre-existing water content gradient, but it is likely to increase the scatter in the final water content.

The second conclusion of our study is that the maximum water content is of the order to 30 \%. Interestingly enough, this value is smaller than assumed water content in solids beyond the iceline, but at the same time similar to the maximum water content in Solar System bodies up to the terrestrial mass range (see Fig. \ref{Fig:Triangular plot}). \rev{In fact, \citet{ormel2017formation} predict moderate water budgets ($\sim$ 10\%) for the TRAPPIST-1 planets when planetary embryos form at the water snowline, migrate to the inner disk and grow  by dry pebbles until a critical planet mass ($\sim$ 0.7 \ME) is reached  that prohibits  further pebble accretion.}  A reduction of water budget for planets that accreted outside of the snowline can include processes such as desiccation of volatile-rich planetesimals by short-lived radionuclides \citep{grimm1993heliocentric,LICHTENBERGepsc}, giant impacts between embryos \citep{genda2005enhanced}, and heating of planets and planetesimals during accretion and collisions which is expected to be more efficient around low-mass stars \citep{lissauer2007planets}. 

In any case, the water budget of the TRAPPIST-1 planets is an important constraint that needs to be fully considered in formation models. As a consequence, better determinations (e.g. by improving the mass, radius, compositional constraints on refractory elements or atmospheric composition) of this quantity is a key in order to make future progress in the understanding of the formation of the system.

%
 %


\section{Conclusions}
\label{conclusion}

The TRAPPIST-1 planets do not follow a single mass-radius trend, but there is some scatter among the bulk densities of planets. Here, we have quantified the origin of this scatter, that is mostly due to different amounts of water, but also to some extent the sizes of rocky interiors and the thicknesses of gas envelopes.

Our analysis characterizes the nature of TRAPPIST-1 planet  interiors while accounting for all available and relevant data. These include the correlated planetary masses \& radii, and stellar irradiation. In addition, we have tested different abundance constraints: a stellar proxy based on stars of similar metallicities as well as a proxy that is based on the most dense and probably a purely rocky planet of the system: planet e. The latter abundance proxy is unique to multi-planetary systems (see Section \ref{proxi}). 
Furthermore, there are data specific to multi-planetary systems that have not been considered in previous studies: the interdependency of planet masses between different planets as derived from TTV analysis, and the interdependency of planet radii between different planets as derived from TTV analysis. Here, we have developed a new resampling scheme (Section \ref{sec:resamp}) that allowed us to incorporate the information on interdependent data (Section \ref{interdep}). \rev{The information that we can gain on the interiors by accounting for interdependent planetary data can be important (up to 20\% differences) and even be as important as an improvement in mass and radius precision (Section \ref{sec:data}).}

\rev{We highlight that the precision on the differential planetary data are much better than on absolute masses and radii. This is because the latter includes stellar uncertainties, while the former does not. By accounting for the correlations among all seven masses, we formally use the knowledge on the differential masses. For multi-planetatary systems, as demonstrated here for TRAPPIST-1, the use of differential planetary data is important for a thorough interior investigation.}

\rev{Systematic biases of data can critically influence interior characterization. Care should be taken with our and all previous interior interpretations that critically depend on planetary masses and densities of TRAPPIST-1. Ongoing observational efforts indicate possible changes in mass accuracies within $\pm$ 24\% \citep{Demoryconf}. In this case, all interiors could be consistent with an increasing water mass fraction with orbital period (within 2-12 \% with respect to 1-$\sigma$ errors) or a uniform water mass fraction of ~7\% (within 1-$\sigma$ errors). 
This is in contrast with our findings based on the most recent data publications \citep{Grimm2018, delrez2018early} that we summarize below:}

\begin{itemize}

\item TRAPPIST-1~e can be a super-Mercury type planet with non-Earth-like bulk abundance. This is obvious from the \rev{high bulk density} as determined by \citep{Grimm2018} and was discussed by \citep{suissa2018trappist}. Here, we have quantified the rocky composition of planet e to be characterized by Fe/Si$_{\rm T1e}= 11.2 \pm 5.7$ and Mg/Si$_{\rm T1e} = 5.7 \pm 3.7$. If the rocky composition of planet e was indeed different from the other planets, it could be due to a giant impact that has not only stripped-off parts of the mantle \citep{benz1988} but also removed volatile-rich layers. Such a scenario would explain why planet e is much drier than the other planets. However, this scenario would require an impactor larger than planet e, which does not exist in the system and it remains to be studied if impacts are a reasonable scenario.  In this scenario, the interiors predicted for all planets might be well described by \U, with the exception of planet e that is best described by \A.
\item Alternatively, TRAPPIST-1~e may not be a super-Mercury type planet. There are two interpretations possible under this premise. 
\begin{itemize}
\item First, it is possible that systematic errors of the TTV analysis due to the limited observation time and yet undetected planets in the system, may bias the planetary masses including planet e. If this is the case, the stellar proxy would be the important constraint to favour rather Earth-like interiors for planet e, dismissing high bulk-density interiors. Consequently, due to the interdependency of planetary data, the information kept in the level of incompatibility of data (in scenario \U using the stellar proxy) propagates to all other planets, yielding better constrained interiors by excluding some interior models. In other words, for all planets there is only a subset of interiors that is in agreement with all available data on all planets, their interdependencies, and the stellar proxy. If this case is true, all interiors are well described by scenario \UCM. 
\item Second, it is possible that the interiors of TRAPPIST-1 planets cannot be described by Earth-like interiors or the suggested stellar proxy. Instead the purely rocky interior of all planets is directly probed by the most dense planet, planet e, assuming that all rocky interiors have similar ratios of rock-forming elements (Mg, Si, Fe). In this case, all predicted interiors are best described by scenario \A.
\end{itemize}
\item Differences between estimated interiors are large when comparing different abundance constraints (\U and \A). This demonstrates the need to better understand the relative amounts of refractory, rock-forming elements in dwarf systems, like TRAPPIST-1, which probably also depend on our knowledge of the age of the system. Unfortunately, direct estimates of the photosphere of the faint TRAPPIST-1 are unavailable. Both possible abundance constraints based on stars with similar metallicities (\U) and based on the most dense planet (\A) can be justified. However it is difficult to state a clear preference.
\item \rev{The information that is kept in the interdependency of planetary data is a valuable constraint that can significantly effect interior estimates. For example, estimated median water budgets can vary up to 20 \% (compare scenarios \U and \UCM). Accounting for interdependency of planetary data compares with changes in data precision of 30\%. }
\item Mass and radius data only carry limited information about planetary interios and additional data types are required to significantly improve interior estimates. For example, the improvement on predicted amounts of water due to more precise data (70\% of nominal data uncertainties) is rather small compared to changes in abundance proxies (\U and \A).
\item  Our inferred ranges of water contents of 0-25\% are high compared to terrestrial Solar System planets, and are smaller by a factor of two compared to predictions from formation studies \citep{alibert2016formation}. Volatile-rich interiors of planets in dwarf star systems are predicted given that the water ice-line is much closer to the star compared to Solar-like system.
\item There is no clear trend of volatile fraction with orbital period. This suggests that either the accreted planetesimals were sufficiently mixed such as to blur otherwise expected increases of water fraction with distance from the star. A corresponding uniform water content is indeed possible within 2-$\sigma$ error bars. Alternatively, migration may have rearranged the planets before they were captured in resonance.
\item Possible delivery of volatiles after formation by cometary impacts \citep{kral2018cometary} of few Earth oceans are tiny compared to our predicted water mass fractions. This means that the overall water budgets were accreted during formation.
Also, the interior degeneracy is large such that uncertainties of predicted water masses are orders of magnitudes larger than possible late-delivered amounts of volatiles. This implies that the data do not allow to validate late delivery of volatiles.
\item The loss of volatiles as predicted by \citep{Bourrier_2017} of several tens of Earth oceans is small compared to the total amount of water that shape the planets, except for planet e. This implies that the ice mass fraction of the bulk accreted material does not significantly exceed 30\%. \rev{If data accuracy changes by up to 20\%, this upper limit could be significantly lower ($\sim$15\%).}
\item The uncertainty in our predicted water mass fractions stems from the degeneracy with the size, structure, and composition of the rocky interior, as well as with the characteristics of the overlying gas envelope. The estimated degeneracy will be generally lower if interior models are employed that only allow for limited variability, e.g., mantles of pure MgSiO$_3$ as employed in \citep{unterborn2017constraining}. Similarly, estimated degeneracies will be larger if interior models are used that allow for interiors that are unlikely to exist in nature, e.g., pure iron cores surrounded by gas envelopes as used in \citep{suissa2018trappist}.
\end{itemize}

\rev{Significant further improvements on our understanding of the TRAPPIST-1 planetary interiors are only expected with higher data accuracy and/or  informative data other than those investigated here, that may include better understanding of their specific host star chemistry, spectroscopic constraints on atmospheric composition (e.g., with JWST, Ariel, E-ELT), or constraints on tidal parameters \citep[e.g.,][]{papaloizou2017trappist}.}

With our study on TRAPPIST-1, we have explored the data types that are specific to multi-planetary systems. Such data will be relevant for the interior characterization of planets in other systems as well. First, there are correlations between the data of different planets that can  carry crucial information for interior characterization. Second, we have demonstrated that it is possible to preferentially probe the rocky interiors of all planets by studying the most dense planet of a multi-planetary system. This study provides new pathways for an improved  interior characterization that is specific to multiplanetary systems.


\begin{acknowledgements}

This work was supported by the Swiss National Foundation under grant PZ00P2\_174028.
It was in part carried out within the frame of the National Centre for Competence in Research PlanetS. 

\end{acknowledgements}

\appendix

\section{Approximation of $\alpha_{\rm max}$}
\label{tlimit}
There is a physical upper limit to the amount of warming by greenhouse gases. The Komabayashi-Ingersoll (KI) limit describes the maximum amount of radiation which can be transferred by a moist atmosphere, which occurs when the transparency $\tau_s$ of the atmosphere becomes very small, i.e., $\tau_s = \tau_{\rm limit}$. 

Here, this limit is represented by $\alpha_{\rm max}$ and that we roughly approximate as follows:

\begin{equation}
\alpha_{\rm max} = \nicefrac{T_{\rm limit} }{ T_{\rm star} \sqrt{\frac{R_{\rm star}}{2 a}}},
\end{equation}
where $R_{\rm star}$ and $T_{\rm star}$ are radius and effective temperature of the host star, $a$ is semi-major axes, and $T_{\rm limit}$ is:

\begin{equation}
\label{tlim}
T_{\rm limit} = \frac{ T_0}{\ln(\frac{\kappa*p_0}{\tau_{\rm limit}g})}.
\end{equation}
Here, $T_0$ is the temperature at some vapor pressure $p_0$ (here, we use $p_0 = 1\times10^{5}$Pa and $T_0 = 373$ K for water, \citep{gold}); $\kappa$ and $\tau_{\rm limit}$ are opacity and optical depth at the KI limit, $g$ is surface gravity. The fraction $\nicefrac{\kappa}{\tau_{\rm limit}}$ is approximated for Earth ($T_{\rm limit} \approx 400$ K) and is estimated to be $10^{-7}$ (in SI units). Thereby, $T_{\rm limit}$ (Eq. \ref{tlim}) scales with the surface gravity. This is a rough estimate for $T_{\rm limit}$ and thus $\alpha_{\rm max}$. More advanced modeling would be required to better determine this limit, but this is outside of the scope of this study.

Equation \ref{tlim} is derived from $\tau_s = \nicefrac{\kappa*p_s}{g}$ and the Clausius-Clapeyron equation, that relates the surface pressure $p_s$ and temperature $T_s$:

\begin{equation}
p_s = p_0 \exp(-\frac{T_0}{T_s}).
\end{equation}

\section{Equivalence between Bayes' theorem and the Tarantola Valette formulation}
\label{BayesTV}

The prior in the Valette formulation is described in data and model space ($\mathcal{D}$ and $\mathcal{M}$).
If the data are \emph{a priori} independent of the model, the prior can be written as:
\begin{equation}
\rho(\dat,\m) = \rho_{d}(\dat) \rho_{m}(\m) \,
\end{equation}
where $ \rho_{d}(\dat)$ and $\rho_{m}(\m)$ are evaluated in data and model space, respectively.
If the data-model mapping $g(\cdot)$ is exact, the forward density $\theta(\dat,\m)$ can be rewritten as:
\begin{equation}
\begin{split}
\theta(\dat,\m) &= \theta(\dat|\m)\theta(\m) \\
                         &=  \theta(\dat|\m)z(\m)  \\
                         & = \delta(\dat-g(\m)) z(\m) \,
\end{split}
\end{equation}
where $\delta$ is the Delta function. If the null information density $z(\dat,\m)$ is constant, we can state the posterior as:
\begin{equation}
\begin{split}
p_m(\m) &= \int_{\mathcal{D}}{p(\dat,\m) d\dat} \\
& = \int_{\mathcal{D}}{ \frac{\rho_{d}(\dat) \rho_{m}(\m)  \delta(\dat-g(\m)) z(\m)}{z(\dat,\m)}}d\dat \\
&=\rho_{d}(g(\m)) p_{m}(\m) \\
&\equiv L_{d}(\m) p_{m}(\m) \,
\end{split}
\end{equation}

where $L_{d}(\m)$  is  the likelihood of the Bayes' formulation.

%


\bibliographystyle{aa}
\bibliography{/Users/cdorn/Documents/libary.bib} 
\label{lastpage}

\end{document}